\documentclass[12pt]{article}

\usepackage{graphicx}
\usepackage{lscape}
\usepackage{citesort}

\begin{document}

\def\ds{\displaystyle}
\def\beq{\begin{equation}}
\def\eeq{\end{equation}}
\def\bea{\begin{eqnarray}}
\def\eea{\end{eqnarray}}
\def\beeq{\begin{eqnarray}}
\def\eeeq{\end{eqnarray}}
\def\ve{\vert}
\def\vel{\left|}
\def\ver{\right|}
\def\nnb{\nonumber}
\def\ga{\left(}
\def\dr{\right)}
\def\aga{\left\{}
\def\adr{\right\}}
\def\lla{\left<}
\def\rra{\right>}
\def\rar{\rightarrow}
\def\nnb{\nonumber}
\def\la{\langle}
\def\ra{\rangle}
\def\ba{\begin{array}}
\def\ea{\end{array}}
\def\tr{\mbox{Tr}}
\def\ssp{{\Sigma^{*+}}}
\def\sso{{\Sigma^{*0}}}
\def\ssm{{\Sigma^{*-}}}
\def\xis0{{\Xi^{*0}}}
\def\xism{{\Xi^{*-}}}
\def\qs{\la \bar s s \ra}
\def\qu{\la \bar u u \ra}
\def\qd{\la \bar d d \ra}
\def\qq{\la \bar q q \ra}
\def\gGgG{\la g^2 G^2 \ra}
\def\q{\gamma_5 \not\!q}
\def\x{\gamma_5 \not\!x}
\def\g5{\gamma_5}
\def\sb{S_Q^{cf}}
\def\sd{S_d^{be}}
\def\su{S_u^{ad}}
\def\ss{S_s^{??}}
\def\sbp{{S}_Q^{'cf}}
\def\sdp{{S}_d^{'be}}
\def\sup{{S}_u^{'ad}}
\def\ssp{{S}_s^{'??}}
\def\sig{\sigma_{\mu \nu} \gamma_5 p^\mu q^\nu}
\def\fo{f_0(\frac{s_0}{M^2})}
\def\ffi{f_1(\frac{s_0}{M^2})}
\def\fii{f_2(\frac{s_0}{M^2})}
\def\O{{\cal O}}
\def\sl{{\Sigma^0 \Lambda}}
\def\es{\!\!\! &=& \!\!\!}
\def\ap{\!\!\! &\approx& \!\!\!}
\def\ar{&+& \!\!\!}
\def\ek{&-& \!\!\!}
\def\kek{\!\!\!&-& \!\!\!}
\def\cp{&\times& \!\!\!}
\def\se{\!\!\! &\simeq& \!\!\!}
\def\eqv{&\equiv& \!\!\!}
\def\kpm{&\pm& \!\!\!}
\def\kmp{&\mp& \!\!\!}


\def\simlt{\stackrel{<}{{}_\sim}}
\def\simgt{\stackrel{>}{{}_\sim}}


\title{
         {\Large
                 {\bf
Double--lepton polarizations in $B \rar \ell^+ \ell^- \gamma$ decay
                 }
         }
      }

\author{\vspace{1cm}\\
{\small T. M. Aliev \thanks
{e-mail: taliev@metu.edu.tr}\,\,,
V. Bashiry
\,\,,
M. Savc{\i} \thanks
{e-mail: savci@metu.edu.tr}} \\
{\small Physics Department, Middle East Technical University,
06531 Ankara, Turkey} }

\date{}

\begin{titlepage}
\maketitle
\thispagestyle{empty}

\begin{abstract}
Double--lepton polarization asymmetries in the $B \rar \ell^+ \ell^-
\gamma$ decay are calculated using the most general, model independent form
of the effective Hamiltonian including all possible forms of the
interaction. The dependencies of the asymmetries on new Wilson coefficients
are investigated. The detectability the averaged double--lepton polarization
asymmetries at LHC is also discussed.  
\end{abstract}
~~~PACS numbers: 12.60.--i, 13.30.--a
\end{titlepage}

\section{Introduction}

Rare radiative leptonic $B_{s(d)} \rar \ell^+ \ell^- \gamma$ decays are 
induced by the flavor--changing neutral current transitions $b\rar s(d)$. 
In the standard model (SM) such processes are described by the penguin and 
box diagrams and have branching ratios $10^{-8}-10^{-15}$ (see for example
\cite{R6801}). These rare decays can not be observed at the running machines
such as Tevatron, BaBar and Belle, but the $B_{s(d)} \rar \mu^+ \mu^-$ and 
$B_{s(d)} \rar \mu^+ \mu^- \gamma$ decays are already studied \cite{R6802}. 
Moreover, the transition form factors and many experimental observables such
as, the branching ratio, photon energy, dilepton mass spectra and charge
asymmetries are calculated for the $B_{s(d)} \rar \ell^+ \ell^- \gamma$ 
decays in \cite{R6803,R6804,R6805,R6806,R6807,R6808,R6809}. 
At the same time $B_{s(d)} \rar \ell^+ \ell^-
\gamma$ decays might be sensitive to the new physics beyond the SM. New
physics effects in these decays can appear in two different ways: either
through the new operators in the effective Hamiltonian which are absent in
the SM, or through new contributions to the Wilson coefficients existing in
the SM. One efficient way for precise determination of the SM parameters and
looking for new physics beyond the SM is studying the lepton polarization
effects. It has been pointed out in \cite{R6810} that some of the single
lepton polarization asymmetries might be too small to be observed and might
not provide sufficient number of observables in checking the structure of
the effective Hamiltonian. In need of more observables, in \cite{R6810},
the maximum number of independent observables have been constructed 
by considering the situation where both lepton polarizations are 
simultaneously measured.

In the present work, we analyze the possibility of searching for new physics
in the $B \rar \ell^+ \ell^- \gamma$ decay by studying the double--lepton
polarization asymmetries, using the most general, model independent form of
the effective Hamiltonian including all possible interactions.  

The work is organized as follows. In section 2, the matrix element for the
$B_s \rar \ell^+ \ell^- \gamma$ is obtained, using the general, model
independent form of the effective Hamiltonian. In section 3, we calculate
the double--lepton polarization asymmetries. Section for is devoted to the
numerical analysis, discussions and conclusions.

\section{Matrix element for the $B_q \rar \ell^+ \ell^- \gamma$ decay}

In this section we derive the matrix element for the $B \rar \ell^+ \ell^-
\gamma$ using the general, model independent form of the effective
Hamiltonian. The matrix element for the process $B \rar \ell^+ \ell^- \gamma$ 
can be obtained from that of the purely leptonic $B \rar \ell^+ \ell^-$ decay. 
At inclusive level the process $B \rar \ell^+ \ell^-$ is described by 
$b \rar q \ell^+ \ell^-$ transition. The effective $b\rar q \ell^+ \ell^-$ 
transition can be written in terms of twelve model independent
four--Fermi interactions in the following form \cite{R6811}:
\bea
\label{e6801}
{\cal H}_{eff} \es \frac{G\alpha}{\sqrt{2} \pi}
 V_{tq}V_{tb}^\ast
\Bigg\{ C_{SL} \, \bar q i \sigma_{\mu\nu} \frac{q^\nu}{q^2}\, L \,b
\, \bar \ell \gamma^\mu \ell + C_{BR}\, \bar q i \sigma_{\mu\nu}
\frac{q^\nu}{q^2} \,R\, b \, \bar \ell \gamma^\mu \ell \nnb \\
\ar C_{LL}^{tot}\, \bar q_L \gamma_\mu b_L \,\bar \ell_L \gamma^\mu \ell_L +
C_{LR}^{tot} \,\bar q_L \gamma_\mu b_L \, \bar \ell_R \gamma^\mu \ell_R +
C_{RL} \,\bar q_R \gamma_\mu b_R \,\bar \ell_L \gamma^\mu \ell_L \nnb \\
\ar C_{RR} \,\bar q_R \gamma_\mu b_R \, \bar \ell_R \gamma^\mu \ell_R +
C_{LRLR} \, \bar q_L b_R \,\bar \ell_L \ell_R +
C_{RLLR} \,\bar q_R b_L \,\bar \ell_L \ell_R
+ C_{LRRL} \,\bar q_L b_R \,\bar \ell_R \ell_L \nnb \\
\ar C_{RLRL} \,\bar q_R b_L \,\bar \ell_R \ell_L+
C_T\, \bar q \sigma_{\mu\nu} b \,\bar \ell \sigma^{\mu\nu}\ell 
+ i C_{TE}\,\epsilon^{\mu\nu\alpha\beta} \bar q \sigma_{\mu\nu} b \,
\bar \ell \sigma_{\alpha\beta} \ell  \Bigg\}~,
\eea
where $C_X$ are the coefficients of the four--Fermi interactions and
\bea
L = \frac{1-\gamma_5}{2} ~,~~~~~~ R = \frac{1+\gamma_5}{2}\nnb~.
\eea
The terms with coefficients $C_{SL}$ and $C_{BR}$ which describe penguin
contributions correspond to $-2 m_s C_7^{eff}$ 
and $-2 m_b C_7^{eff}$ in the SM, respectively. The next four terms in this
expression are the vector interactions. The interaction terms containing
$C_{LL}^{tot}$ and $C_{LR}^{tot}$ in the SM have the form
$C_9^{eff}-C_{10}$ and $C_9^{eff}+C_{10}$, respectively. Inspired by this 
$C_{LL}^{tot}$ and $C_{LR}^{tot}$ will be written as
\bea
C_{LL}^{tot} \es C_9^{eff} - C_{10} + C_{LL}~, \nnb \\
C_{LR}^{tot} \es C_9^{eff} + C_{10} + C_{LR}~, \nnb
\eea
where $C_{LL}$ and $C_{LR}$ describe contributions from new physics.
The terms with coefficients $C_{LRLR}$, $C_{RLLR}$, $C_{LRRL}$ and 
$C_{RLRL}$ describe the scalar type interactions. The last two 
terms in Eq. (\ref{e6801}) with the coefficients $C_T$ and $C_{TE}$
describe the tensor type interactions.

Having presented the general form of the effective Hamiltonian the next
problem is the calculation of the matrix element of the $B_q \rar \ell^+ \ell^-
\gamma$ decay. This matrix element can be written as the sum of the
structure--dependent and inner Bremsstrahlung parts 
\bea
\label{e6802}
{\cal M} = {\cal M}_{SD}+{\cal M}_{IB}~. 
\eea
The matrix element for the structure--dependent part ${\cal M}_{SD}$ can be
obtained by sandwiching the effective Hamiltonian between initial $B$ and
final photon states, i.e., $\lla \gamma\vel{\cal H}_{eff}\ver B \rra$.
It follows from Eq. (\ref{e6801}) that in order to calculate 
${\cal M}_{SD}$, the following matrix elements are needed
\bea
\label{e6803}
&&\lla \gamma\vel \bar s \gamma_\mu (1 \mp \gamma_5)
b \ver B \rra~,\nnb \\
&&\lla \gamma \vel \bar s \sigma_{\mu\nu} b \ver B \rra~, \nnb \\
&&\lla \gamma \vel \bar s (1 \mp \gamma_5) b
\ver B \rra~.
\eea
The first two of the matrix elements in Eq. (\ref{e6803}) are defined as
\cite{R6803,R6812,R6813}
\bea
\label{e6804}
\lla \gamma(k) \vel \bar q \gamma_\mu
(1 \mp \gamma_5) b \ver B(p_B) \rra \es
\frac{e}{m_B^2} \Big\{
\epsilon_{\mu\nu\lambda\sigma} \varepsilon^{\ast\nu} q^\lambda
k^\sigma g(q^2) \pm i\,
\Big[ \varepsilon^{\ast\mu} (k q) -
(\varepsilon^\ast q) k^\mu \Big] f(q^2) \Big\}~,\\ \nnb \\
\label{e6805}
\lla \gamma(k) \vel \bar q \sigma_{\mu\nu} b \ver B(p_B) \rra \es 
\frac{e}{m_B^2}
\epsilon_{\mu\nu\lambda\sigma} \Big[
G \varepsilon^{\ast\lambda} k^\sigma +
H \varepsilon^{\ast\lambda} q^\sigma +
N (\varepsilon^\ast q) q^\lambda k^\sigma \Big]~,
\eea
respectively, where $\varepsilon^\ast$ and $k$ are the four vector polarization
and momentum of the photon, respectively, $q$ is the momentum transfer,
$p_B$ is the momentum of the $B$ meson and $g(q^2)$, $f(q^2)$, $G(q^2)$, 
$H(q^2)$ and $N(q^2)$ are the $B \rar \gamma$ transition form factors. 
The matrix element 
$\lla \gamma(k) \vel \bar s \sigma_{\mu\nu} \gamma_5 b \ver B(p_B) \rra$
can be obtained from Eq. (\ref{e6805}) using the identity
\bea
\sigma_{\mu\nu} = - \frac{i}{2}\epsilon_{\mu\nu\alpha\beta}
\sigma^{\alpha\beta} \gamma_5~.\nnb
\eea
The matrix elements 
$\lla \gamma(k) \vel \bar s (1 \mp \gamma_5) b \ver B(p_B) \rra$ 
and
$\lla \gamma \vel \bar s i \sigma_{\mu\nu} q^\nu b \ver B \rra$ can be
calculated by contracting both sides of the  Eqs. (\ref{e6804}) and (\ref{e6805})
with $q^\mu$ and $q^\nu$, respectively. We get then
\bea
\label{e6806}
\lla \gamma(k) \vel \bar s (1 \mp \gamma_5) b \ver B(p_B) \rra
\es 0~, \\
\label{e6807}
\lla \gamma \vel \bar s i \sigma_{\mu\nu} q^\nu b \ver B \rra \es
\frac{e}{m_B^2} i\, \epsilon_{\mu\nu\alpha\beta} q^\nu
\varepsilon^{\alpha\ast} k^\beta G~.
\eea
The matrix element $\lla \gamma
\vel \bar s i \sigma_{\mu\nu} q^\nu (1+\gamma_5) b \ver B \rra$ can be
written in terms of the form factors that are calculated in framework of the QCD
sum rules \cite{R6803} and light front model \cite{R6804} as follows 
\bea
\label{e6808}
\lla \gamma \vel \bar s i \sigma_{\mu\nu} q^\nu (1+\gamma_5) b \ver B \rra \es
\frac{e}{m_B^2} \Big\{
\epsilon_{\mu\alpha\beta\sigma} \, \varepsilon^{\alpha\ast} q^\beta k^\sigma
g_1(q^2)
+ i\,\Big[\varepsilon_\mu^\ast (q k) - (\varepsilon^\ast q) k_\mu \Big]
f_1(q^2) \Big\}~.
\eea
It should be noted that these form factors were calculated in framework of
the light--front model in \cite{R6813}.
Eqs. (\ref{e6805}), (\ref{e6807}) and (\ref{e6808}) allow us to
express $G,~H$ and $N$ in terms of the form factors $g_1$ and $f_1$. Using
Eqs. (\ref{e6804})--(\ref{e6808}), ${\cal M}_{SD} $ can be expressed as
\bea
\label{e6809}
{\cal M}_{SD} \es \frac{\alpha G_F}{4 \sqrt{2} \, \pi} V_{tb} V_{tq}^* 
\frac{e}{m_B^2} \,\Bigg\{
\bar \ell \gamma^\mu (1-\gamma_5) \ell \, \Big[
A_1 \epsilon_{\mu \nu \alpha \beta} 
\varepsilon^{\ast\nu} q^\alpha k^\beta + 
i \, A_2 \Big( \varepsilon_\mu^\ast (k q) - 
(\varepsilon^\ast q ) k_\mu \Big) \Big] \nnb \\
\ar \bar \ell \gamma^\mu (1+\gamma_5) \ell \, \Big[
B_1 \epsilon_{\mu \nu \alpha \beta} 
\varepsilon^{\ast\nu} q^\alpha k^\beta 
+ i \, B_2 \Big( \varepsilon_\mu^\ast (k q) - 
(\varepsilon^\ast q ) k_\mu \Big) \Big] \nnb \\
\ar i \, \epsilon_{\mu \nu \alpha \beta} 
\bar \ell \sigma^{\mu\nu}\ell \, \Big[ G \varepsilon^{\ast\alpha} k^\beta 
+ H \varepsilon^{\ast\alpha} q^\beta + 
N (\varepsilon^\ast q) q^\alpha k^\beta \Big] \\
\ar i \,\bar \ell \sigma_{\mu\nu}\ell \, \Big[
G_1 (\varepsilon^{\ast\mu} k^\nu - \varepsilon^{\ast\nu} k^\mu) + 
H_1 (\varepsilon^{\ast\mu} q^\nu - \varepsilon^{\ast\nu} q^\mu) +
N_1 (\varepsilon^\ast q) (q^\mu k^\nu - q^\nu k^\mu) \Big] \Bigg\}~,\nnb
\eea
where
\bea
\label{e6810}
A_1 \es \frac{1}{q^2} \Big( C_{BR} + C_{SL} \Big) g_1 +
\Big( C_{LL}^{tot} + C_{RL} \Big) g ~, \nnb \\
A_2 \es \frac{1}{q^2} \Big( C_{BR} - C_{SL} \Big) f_1 +
\Big( C_{LL}^{tot} - C_{RL} \Big) f ~, \nnb \\
B_1 \es \frac{1}{q^2} \Big( C_{BR} + C_{SL} \Big) g_1 +
\Big( C_{LR}^{tot} + C_{RR} \Big) g ~, \nnb \\
B_2 \es \frac{1}{q^2} \Big( C_{BR} - C_{SL} \Big) f_1 +
\Big( C_{LR}^{tot} - C_{RR} \Big) f ~, \nnb \\
G \es 4 C_T g_1 ~, \nnb \\
N \es - 4 C_T \frac{1}{q^2} (f_1+g_1) ~, \\
H \es N (qk) ~, \nnb \\
G_1 \es - 8 C_{TE} g_1 ~, \nnb \\
N_1 \es 8 C_{TE} \frac{1}{q^2} (f_1+g_1) ~, \nnb \\ 
H_1 \es N_1(qk)~. \nnb
\eea
For the inner Bremsstrahlung part we get
\bea
\label{e6811}
{\cal M}_{IB} \es \frac{\alpha G_F}{4 \sqrt{2} \, \pi} V_{tb} V_{tq}^*  
e f_B i \,\Bigg\{
F\, \bar \ell  \Bigg(
\frac{{\not\!\varepsilon}^\ast {\not\!p}_B}{2 p_1 k} - 
\frac{{\not\!p}_B {\not\!\varepsilon}^\ast}{2 p_2 k} \Bigg) 
\gamma_5 \ell \nnb \\
\ar F_1 \, \bar \ell  \Bigg[
\frac{{\not\!\varepsilon}^\ast {\not\!p}_B}{2 p_1 k} -
\frac{{\not\!p}_B {\not\!\varepsilon}^\ast}{2 p_2 k} +
2 m_\ell \Bigg(\frac{1}{2 p_1 k} + \frac{1}{2 p_2 k}\Bigg)
{\not\!\varepsilon}^\ast \Bigg] \ell \Bigg\}~,
\eea
where we have used
\bea
\la 0 \ve \bar s \gamma_\mu \gamma_5 b \ve B \ra \es 
-~i f_B p_{B\mu}~, \nnb \\
\la 0 \ve \bar s \sigma_{\mu\nu} (1+\gamma_5) b \ve B \ra \es 0~,\nnb
\eea
and conservation of the vector current.
The functions $F$ and $F_1$ are defined as follows
\bea
\label{e6812}
F \es 2 m_\ell \Big( C_{LR}^{tot} - C_{LL}^{tot} + C_{RL} - C_{RR} \Big)
+ \frac{m_B^2}{m_b}
\Big( C_{LRLR} - C_{RLLR} - C_{LRRL} + C_{RLRL} \Big)~, \nnb \\
F_1 \es\frac{m_B^2}{m_b} \Big( C_{LRLR} - C_{RLLR} + C_{LRRL} - C_{RLRL}
\Big)~.
\eea

\section{Double--lepton polarization asymmetries in $B_q \rar \ell^+ \ell^-
\gamma$ decay}

In the present section we calculate the double--lepton polarization
asymmetries, i.e., when polarizations of both leptons are taken into
account. In order to calculate the double lepton polarization asymmetries we
define the following orthogonal unit vectors $s_i^\pm$ in the rest frame of
$\ell^\pm$
\bea
\label{e6813}
s^{-\mu}_L \es \ga 0,\vec{e}_L^{\,-}\dr =
\ga 0,\frac{\vec{p}_-}{\vel\vec{p}_- \ver}\dr~, \nnb \\
s^{-\mu}_N \es \ga 0,\vec{e}_N^{\,-}\dr = \ga 0,\frac{\vec{p}_\Lambda\times
\vec{p}_-}{\vel \vec{p}_\Lambda\times \vec{p}_- \ver}\dr~, \nnb \\
s^{-\mu}_T \es \ga 0,\vec{e}_T^{\,-}\dr = \ga 0,\vec{e}_N^{\,-}
\times \vec{e}_L^{\,-} \dr~, \nnb \\
s^{+\mu}_L \es \ga 0,\vec{e}_L^{\,+}\dr =
\ga 0,\frac{\vec{p}_+}{\vel\vec{p}_+ \ver}\dr~, \nnb \\ 
s^{+\mu}_N \es \ga 0,\vec{e}_N^{\,+}\dr = \ga 0,\frac{\vec{p}_\Lambda\times
\vec{p}_+}{\vel \vec{p}_\Lambda\times \vec{p}_+ \ver}\dr~, \nnb \\
s^{+\mu}_T \es \ga 0,\vec{e}_T^{\,+}\dr = \ga 0,\vec{e}_N^{\,+}
\times \vec{e}_L^{\,+}\dr~,
\eea
where $\vec{p}_\pm$ and $\vec{k}$ are the three--momenta of the
leptons $\ell^\pm$ and photon in the
center of mass frame (CM) of $\ell^- \,\ell^+$ system, respectively.
Transformation of unit vectors from the rest frame of the leptons to CM
frame of leptons can be accomplished by the Lorentz boost. Boosting of the
longitudinal unit vectors $s_L^{\pm\mu}$ yields
\bea
\label{e6814}
\ga s^{\mp\mu}_L \dr_{CM} \es \ga \frac{\vel\vec{p}_\mp \ver}{m_\ell}~,
\frac{E_\ell \vec{p}_\mp}{m_\ell \vel\vec{p}_\mp \ver}\dr~,
\eea
where $\vec{p}_+ = - \vec{p}_-$, $E_\ell$ is the energy of the lepton in the
CM frame and $m_\ell$ is its mass.
The unit vectors $s_N^{\pm\mu}$, $s_T^{\pm\mu}$ are unchanged
under Lorentz transformation.

Having obtained necessary expressions, we now define the
double--polarization asymmetries as follows \cite{R6810}:

\bea
\label{e6618}
P_{ij}(q^2) \es
\frac{
\Big( \ds \frac{d\Gamma(\vec{s}^{\,-}_i,\vec{s}^{\,+}_j)}{dq^2} -
      \ds \frac{d\Gamma(-\vec{s}^{\,-}_i,\vec{s}^{\,+}_j)}{dq^2} \Big) -
\Big( \ds \frac{d\Gamma(\vec{s}^{\,-}_i,-\vec{s}^{\,+}_j)}{dq^2} -
      \ds \frac{d\Gamma(-\vec{s}^{\,-}_i,-\vec{s}^{\,+}_j)}{dq^2} \Big)
     }
     {
\Big( \ds \frac{d\Gamma(\vec{s}^{\,-}_i,\vec{s}^{\,+}_j)}{dq^2} +
      \ds \frac{d\Gamma(-\vec{s}^{\,-}_i,\vec{s}^{\,+}_j)}{dq^2} \Big) +
\Big( \ds \frac{d\Gamma(\vec{s}^{\,-}_i,-\vec{s}^{\,+}_j)}{dq^2} +
      \ds \frac{d\Gamma(-\vec{s}^{\,-}_i,-\vec{s}^{\,+}_j)}{dq^2} \Big)
     }~,    
\eea
where, the first subindex $i$ represents lepton and the second one
antilepton. Using this definition of $P_{ij}$, and after lengthy
calculations, for the nine double--lepton polarization asymmetries
we get

\bea
\label{e6816}
P_{LL} \es 
\frac{1}{\Delta} \Bigg\{
- \frac{4}{\hat{m}_\ell} m_B \hat{s}^2 (1- \hat{s}) (1-v^2) 
\mbox{\rm Im}[(A_2^\ast + B_2^\ast) H_1] \nnb \\
\ar 16 \hat{s} (1- \hat{s}) \mbox{\rm Re}[v^2 G^\ast H -
(1-2 v^2) G_1^\ast H_1] \nnb \\
\ar 16 \hat{s}^2 \Big[ v^2 \vel H \ver^2 - 
(1-2 v^2) \vel H_1 \ver^2 \Big] \nnb \\
\ar \frac{1}{2} f_B^2 m_B^4 \Big\{ (1-\hat{s})^2 ({\cal I}_1+{\cal I}_4) - [2 \hat{s} + 
(1+\hat{s}^2) v^2] {\cal I}_3 + [2 \hat{s} - (1+\hat{s}^2) v^2] {\cal I}_6 \Big\} 
\vel F \ver^2 \nnb \\
\ar \frac{1}{2} f_B^2 m_B^4 \Big\{ - (1-\hat{s})^2 {\cal I}_1 +      
v^2 [1 + \hat{s} (2 - \hat{s} + 2\hat{s} v^2)] {\cal I}_3 \nnb \\  
\ek (1-\hat{s}) [(1-\hat{s}) {\cal I}_4 + 2 \hat{s} v (1-v^2) {\cal I}_5] +
v^2 [1 - \hat{s} (2 + \hat{s} - 2\hat{s} v^2)] {\cal I}_6
\Big\} \vel F_1 \ver^2 \nnb \\
\ek 4 f_B m_B^2 \hat{s} v [(1-\hat{s}) v {\cal I}_{8} + (1+\hat{s}) {\cal I}_{9}] 
\mbox{\rm Re} [F^\ast H] \nnb \\
\ek 4 f_B m_B^2 \hat{s} [(1-\hat{s}) v^2 {\cal I}_{8} - (1-\hat{s}-2 v^2) {\cal I}_{9}] 
\mbox{\rm Im} [F_1^\ast H_1] \nnb \\
\ar f_B m_B^2 \hat{s} \Big[                  
8 (1- \hat{s} + 2 \hat{s} v^2) + m_B^2 (1-\hat{s})
(4+v^2-4 \hat{s} + 3 \hat{s} v^2 + 2 \hat{s} v^4) {\cal I}_{8} \nnb \\
\ar m_B^2 (1-\hat{s}) (3-3 \hat{s} - 2 v^2 + 4 \hat{s} v^2) {\cal I}_{9}]
\Big] \mbox{\rm Im} [F_1^\ast N_1] \nnb \\
\ar f_B \Big[                         
8 (1- \hat{s} - 2 \hat{s} v^2) + m_B^2 (1-\hat{s})  
(4+v^2-4 \hat{s}+3 \hat{s} v^2 -2 \hat{s} v^4) {\cal I}_{8} \nnb \\
\ek m_B^2 (1-\hat{s}) (1- \hat{s} + 4 v^2 - 2 \hat{s} v^2) {\cal I}_{9}]
\Big] \mbox{\rm Im} [F_1^\ast G_1] \nnb \\
\ar f_B \Big[ 
- 8 (1+ \hat{s}) + m_B^2 (1-\hat{s})  
(4-v^2 -4 \hat{s} + 3 \hat{s} v^2) {\cal I}_{8} +
m_B^2 (1-\hat{s}) (1+ \hat{s} - 4 v^2) {\cal I}_{9}]
\Big] \mbox{\rm Re} [F^\ast G] \nnb \\
\ar f_B m_B^2 \hat{s} \Big[                                
-8 (1+ \hat{s}) + 
m_B^2 (1-\hat{s}) (3 - \hat{s}) v^2 {\cal I}_{8} +
m_B^2 (1- \hat{s}^2) {\cal I}_{9}         
\Big] \mbox{\rm Re} [F^\ast N] \nnb \\
\ar \frac{1}{4 \hat{m}_\ell} f_B m_B \Big[ 
4 \hat{s} (1+v^2 -\hat{s} + 3 \hat{s} v^2) +
m_B^2 \hat{s} (1-\hat{s}) (4 -4 \hat{s} - 3 v^2 + 7 \hat{s} v^2 + v^4
-\hat{s} v^4) {\cal I}_{8} \nnb \\
\ek m_B^2 \hat{s} (1-\hat{s}) (1- \hat{s} + v^2 + 7 \hat{s} v^2 - 
4 \hat{s} v^4) {\cal I}_{9}] \Big]
\mbox{\rm Re} [(A_2^\ast + B_2^\ast) F_1] \nnb \\
\ek \frac{1}{2 \hat{m}_\ell} f_B m_B \hat{s} \Big[ 
8 (1+\hat{s}) v^2 +
m_B^2 (1-\hat{s}) (2-2 \hat{s} -2 v^2 + 2 \hat{s} v^2 + 
v^4 +\hat{s} v^4) {\cal I}_{8} \nnb \\
\ek m_B^2 (1-\hat{s}^2) v^2 {\cal I}_{9}] \Big] 
\mbox{\rm Re} [(A_1^\ast + B_1^\ast) F] \nnb \\
\ek \frac{8}{3} (1-\hat{s})^2 (1-3 v^2) \Big( \vel G \ver^2 +
\vel G_1 \ver^2 \Big) \nnb \\
\ar \frac{2}{3 \hat{m}_\ell} m_B \hat{s} (1-\hat{s})^2 (1-v^2)
\Big\{ m_B^2 \hat{s} \mbox{\rm Im} [(A_2^\ast + B_2^\ast) N_1]
- 2 \mbox{\rm Im} [(A_2^\ast + B_2^\ast) G_1] \Big\}\nnb \\
\ar \frac{4}{3 \hat{m}_\ell} m_B \hat{s} (1-\hat{s})^2 (1-v^2)
\mbox{\rm Re} [(A_1^\ast + B_1^\ast) G] \nnb \\
\ek \frac{1}{3 \hat{m}_\ell^2} m_B^2 \hat{s}^2 (1-\hat{s})^2 (1-v^2)^2
\mbox{\rm Re} [A_1^\ast B_1 + A_2^\ast B_2] \nnb \\
\ek \frac{2}{3} m_B^2 \hat{s} (1-\hat{s})^2 (1+3 v^2)  
\Big( \vel A_1 \ver^2 + \vel A_2 \ver^2 + 
\vel B_1 \ver^2 + \vel B_2 \ver^2 \Big) \nnb \\
\ar \frac{8}{3} m_B^2 \hat{s} (1-\hat{s})^2            
\Big[ (1-2 v^2) \mbox{\rm Re} [G_1^\ast N_1]                             
- v^2 \mbox{\rm Re} [G^\ast N] \Big] \nnb \\
\ek \frac{4}{3} m_B^4 \hat{s}^2 (1-\hat{s})^2            
\Big[ v^2 \vel N \ver^2 - (1-2 v^2) \vel N_1 \ver^2 \Big] 
\Bigg\}~, \\ \nnb \\
\label{e6817}
P_{LN} \es
\frac{1}{\Delta} \Bigg\{
2 f_B^2 m_B^4 \hat{m}_\ell \sqrt{\hat{s}} (1-\hat{s}) v^2 
\mbox{\rm Im} [F_1 F^\ast] ({\cal I}_2+{\cal I}_4) \nnb \\
\ar f_B m_B^3 \sqrt{\hat{s}} (1-\hat{s}^2) v^2 
\mbox{\rm Im} [A_1^\ast (F_1 + F) + B_1^\ast (F_1 - F)] {\cal I}_7 \nnb \\
\ar 8 f_B m_B^2 \hat{m}_\ell \sqrt{\hat{s}} (1-\hat{s}) v^2
\mbox{\rm Im} [F_1^\ast G] {\cal I}_7 \nnb \\
\ar \pi m_B \sqrt{\hat{s}} (1-\hat{s})^2 v^2
\mbox{\rm Re} [ (A_1^\ast - A_2^\ast + B_1^\ast +B_2^\ast) G_1] \nnb \\
\ek \pi m_B \sqrt{\hat{s}} (1-\hat{s})^2 v^2 
\mbox{\rm Im} [ (A_1^\ast - A_2^\ast - B_1^\ast - B_2^\ast) G] \nnb \\
\ar 2 \pi m_B \sqrt{\hat{s}^3} (1-\hat{s}) v^2
\Big[ \mbox{\rm Re} [ (A_1^\ast + B_1^\ast) H_1] -   
\mbox{\rm Im} [ (A_1^\ast - B_1^\ast) H] \nnb \\
\ek 4 \pi f_B m_B \sqrt{\hat{s}} (1-\hat{s}) (1-\sqrt{1-v^2})
\mbox{\rm Im} [ (A_2^\ast - B_2^\ast) F_1 + (A_2^\ast + B_2^\ast) F] 
\Bigg\}~, \\ \nnb \\
\label{e6818}
P_{NL} \es
\frac{1}{\Delta} \Bigg\{
2 f_B^2 m_B^4 \hat{m}_\ell \sqrt{\hat{s}} (1-\hat{s}) v^2 
\mbox{\rm Im} [F_1 F^\ast] ({\cal I}_2+{\cal I}_4) \nnb \\
\ar f_B m_B^3 \sqrt{\hat{s}} (1-\hat{s}^2) v^2 
\mbox{\rm Im} [A_1^\ast (F_1 - F) + B_1^\ast (F_1 + F)] {\cal I}_7 \nnb \\
\ar 8 f_B m_B^2 \hat{m}_\ell \sqrt{\hat{s}} (1-\hat{s}) v^2
\mbox{\rm Im} [F_1^\ast G] {\cal I}_7 \nnb \\
\ar \pi m_B \sqrt{\hat{s}} (1-\hat{s})^2 v^2
\mbox{\rm Re} [ (A_1^\ast + A_2^\ast + B_1^\ast - B_2^\ast) G_1] \nnb \\
\ar \pi m_B \sqrt{\hat{s}} (1-\hat{s})^2 v^2
\mbox{\rm Im} [ (A_1^\ast + A_2^\ast - B_1^\ast + B_2^\ast) G] \nnb \\
\ar 2 \pi m_B \sqrt{\hat{s}^3} (1-\hat{s}) v^2
\Big[ \mbox{\rm Re} [ (A_1^\ast + B_1^\ast) H_1] +
\mbox{\rm Im} [ (A_1^\ast - B_1^\ast) H] \Big] \nnb \\
\ar 4 \pi f_B m_B \sqrt{\hat{s}} (1-\hat{s}) (1-\sqrt{1-v^2})
\mbox{\rm Im} [ (A_2^\ast - B_2^\ast) F_1 - (A_2^\ast + B_2^\ast) F]
\Bigg\}~, \\ \nnb \\
\label{e6819}
P_{LT} \es
\frac{1}{\Delta} \Bigg\{
- \frac{1}{\sqrt{\hat{s}}} f_B^2 m_B^4 \hat{m}_\ell (1-\hat{s}) v
\Big[ (1-\hat{s}) \vel F_1 \ver^2 + (1+\hat{s}) \vel F \ver^2 \Big]
({\cal I}_2+{\cal I}_4) \nnb \\
\ar 8 f_B m_B^2 \hat{m}_\ell \sqrt{\hat{s}} (1-\hat{s}) v 
\Big(\mbox{\rm Im} [F_1^\ast H_1] + \mbox{\rm Re} [F^\ast H] \Big) 
{\cal I}_7 \nnb \\
\ar f_B m_B^3 \sqrt{\hat{s}} (1-\hat{s})^2 v
\mbox{\rm Re} [(A_2^\ast + B_2^\ast) F_1] {\cal I}_7 \nnb \\
\ar \frac{4}{v} \pi f_B m_B \sqrt{\hat{s}} (1-\hat{s}) (1-\sqrt{1-v^2})
\mbox{\rm Re}[ (A_2^\ast - B_2^\ast) F] \nnb \\
\ek \frac{4}{v} \pi f_B m_B \sqrt{\hat{s}} [1-\hat{s} (1-2 v^2)]
(1-\sqrt{1-v^2})
\mbox{\rm Re}[ (A_1^\ast - B_1^\ast) F_1] \nnb \\
\ek \frac{4}{\sqrt{\hat{s}}} \pi \hat{m}_\ell (1-\hat{s})^2 v
\Big( \vel G_1 \ver^2 + \vel G_1 \ver^2 \Big) \nnb \\
\ek 8 \pi \hat{m}_\ell \sqrt{\hat{s}} (1-\hat{s}) v
\mbox{\rm Re}[ G_1^\ast H_1 + G^\ast H] \nnb \\
\ar \pi m_B \sqrt{\hat{s}} (1-\hat{s})^2 v
\mbox{\rm Im}[ (A_1^\ast - A_2^\ast - B_1^\ast - B_2^\ast) G_1] \nnb \\
\ar \pi m_B \sqrt{\hat{s}} (1-\hat{s})^2 v \Big[
\mbox{\rm Re}[ (A_1^\ast - A_2^\ast + B_1^\ast + B_2^\ast) G]
+ 2 m_B \hat{m}_\ell 
\mbox{\rm Re}[A_1^\ast A_2 - B_1^\ast B_2] \Big] \nnb \\
\ar 2 \pi m_B \sqrt{\hat{s}^3} (1-\hat{s}) v
\Big[ \mbox{\rm Im}[(A_1^\ast - B_1^\ast) H_1] +
\mbox{\rm Re}[(A_1^\ast + B_1^\ast) H] \Big] \nnb \\
\ek \frac{4}{v} \pi f_B m_B \sqrt{\hat{s}} (1+\hat{s}) (1-\sqrt{1-v^2})
\mbox{\rm Re}[(A_1^\ast + B_1^\ast) F] \Big] \nnb \\
\ar \frac{32}{\sqrt{\hat{s}} v} \pi \hat{m}_\ell (1-\hat{s})
(1-\sqrt{1-v^2})
\mbox{\rm Im}[F_1^\ast G_1] \nnb \\
\ar \frac{32}{\sqrt{\hat{s}} v} \pi \hat{m}_\ell (1-\sqrt{1-v^2})
\mbox{\rm Re}[F^\ast G]
\Bigg\}~, \\ \nnb \\
\label{e6820}
P_{TL} \es
\frac{1}{\Delta} \Bigg\{
- \frac{1}{\sqrt{\hat{s}}} f_B^2 m_B^4 \hat{m}_\ell (1-\hat{s}) v
\Big[ (1-\hat{s}) \vel F_1 \ver^2 + (1+\hat{s}) \vel F \ver^2 \Big]
({\cal I}_2+{\cal I}_4) \nnb \\
\ar 8 f_B m_B^2 \hat{m}_\ell \sqrt{\hat{s}} (1-\hat{s}) v 
\Big(\mbox{\rm Im} [F_1^\ast H_1] + \mbox{\rm Re} [F^\ast H] \Big) 
{\cal I}_7 \nnb \\
\ar f_B m_B^3 \sqrt{\hat{s}} (1-\hat{s})^2 v
\mbox{\rm Re} [(A_2^\ast + B_2^\ast) F_1] {\cal I}_7 \nnb \\
\ek \frac{4}{v} \pi f_B m_B \sqrt{\hat{s}} (1-\hat{s}) (1-\sqrt{1-v^2})
\mbox{\rm Re}[ (A_2^\ast - B_2^\ast) F] \nnb \\
\ar \frac{4}{v} \pi f_B m_B \sqrt{\hat{s}} [1-\hat{s} (1-2 v^2)]
(1-\sqrt{1-v^2})
\mbox{\rm Re}[ (A_1^\ast - B_1^\ast) F_1] \nnb \\
\ek \frac{4}{\sqrt{\hat{s}}} \pi \hat{m}_\ell (1-\hat{s})^2 v
\Big( \vel G_1 \ver^2 + \vel G_1 \ver^2 \Big) \nnb \\
\ek 8 \pi \hat{m}_\ell \sqrt{\hat{s}} (1-\hat{s}) v
\mbox{\rm Re}[ G_1^\ast H_1 + G^\ast H] \nnb \\
\ek \pi m_B \sqrt{\hat{s}} (1-\hat{s})^2 v
\mbox{\rm Im}[ (A_1^\ast + A_2^\ast - B_1^\ast + B_2^\ast) G_1] \nnb \\
\ar \pi m_B \sqrt{\hat{s}} (1-\hat{s})^2 v \Big[
\mbox{\rm Re}[ (A_1^\ast + A_2^\ast + B_1^\ast - B_2^\ast) G]
- 2 m_B \hat{m}_\ell 
\mbox{\rm Re}[A_1^\ast A_2 - B_1^\ast B_2] \Big] \nnb \\
\ek 2 \pi m_B \sqrt{\hat{s}^3} (1-\hat{s}) v
\Big[ \mbox{\rm Im}[(A_1^\ast - B_1^\ast) H_1] -
\mbox{\rm Re}[(A_1^\ast + B_1^\ast) H] \Big] \nnb \\
\ek \frac{4}{v} \pi f_B m_B \sqrt{\hat{s}} (1+\hat{s}) (1-\sqrt{1-v^2})
\mbox{\rm Re}[(A_1^\ast + B_1^\ast) F] \Big] \nnb \\
\ar \frac{32}{\sqrt{\hat{s}} v} \pi \hat{m}_\ell (1-\hat{s})
(1-\sqrt{1-v^2})
\mbox{\rm Im}[F_1^\ast G_1] \nnb \\
\ar \frac{32}{\sqrt{\hat{s}} v} \pi \hat{m}_\ell (1-\sqrt{1-v^2})
\mbox{\rm Re}[F^\ast G]
\Bigg\}~, \\ \nnb \\
\label{e6821}
P_{NT} \es
\frac{1}{\Delta} \Bigg\{
-16 \hat{s} (1-\hat{s}) v \mbox{\rm Re}[G^\ast H_1 + G_1^\ast H] \nnb \\
\ek 16 m_B \hat{m}_\ell \hat{s} (1-\hat{s}) v 
\mbox{\rm Im}[(A_2^\ast + B_2^\ast) H] \nnb \\
\ar \frac{16}{v^2} f_B \hat{s} \bigg[ 2 v - (1-v^2) 
\ln \bigg( \frac{1+v}{1-v}\bigg) \bigg]
\mbox{\rm Re}[F^\ast G_1] \nnb \\
\ar 2 f_B m_B^3 \hat{m}_\ell (1-\hat{s})^2 v
\mbox{\rm Im}[ A_1^\ast (F_1-F) + B_1^\ast (F_1+F)]
({\cal I}_{8}-{\cal I}_{9}) \nnb \\
\ek 32 \hat{s}^2 v \mbox{\rm Im}[ H_1^\ast H] \nnb \\
\ek f_B^2 m_B^4 \hat{s} \mbox{\rm Im}[ F_1^\ast F]
\Big\{ v [ 3 + v^2 - \hat{s} (1-v^2)] {\cal I}_3 -
(1-v^2) [ (1-\hat{s}) {\cal I}_5 + (1+\hat{s}) {\cal I}_6 ] \Big\} \nnb \\
\ek 4 f_B m_B^4 (1-\hat{s}) [\hat{s} - 2 \hat{m}_\ell^2 (1+\hat{s})] v
\mbox{\rm Im}[ F_1^\ast N] {\cal I}_{8} \nnb \\
\ar 16 f_B \hat{s} \bigg[ 2 v - (1-v^2) \ln \bigg( \frac{1+v}{1-v}\bigg)
\bigg] \mbox{\rm Im}[F_1^\ast G] \nnb \\
\ek 2 f_B m_B^3 \hat{m}_\ell (1-\hat{s}^2) v 
\mbox{\rm Im}[(A_2^\ast+B_2^\ast) F] ({\cal I}_{8}-{\cal I}_{9}) \nnb \\
\ar 4 f_B m_B^2 \hat{s} v \mbox{\rm Re}[F^\ast H_1]  
\Big[ (1-\hat{s}) {\cal I}_{8} + (1+\hat{s}) {\cal I}_{9} \Big] \nnb \\
\ar 4 f_B m_B^2 \hat{s} v \mbox{\rm Im}[F_1^\ast H]
\Big[ (1-\hat{s}) v^2 {\cal I}_{8} + (1-\hat{s} + 2 \hat{s} v^2) 
{\cal I}_{9}\Big] \nnb \\
\ar \frac{16}{v^2} f_B m_B^2 \hat{s} \bigg[2 v - \ln \bigg( \frac{1+v}{1-v}
\bigg) \bigg] \mbox{\rm Re}[F^\ast N_1] \nnb \\
\ek \frac{8}{v^2} f_B m_B \hat{m}_\ell (1-\hat{s})
\bigg[2 v - (1-v^2) \ln \bigg( \frac{1+v}{1-v} \bigg) \bigg]
\mbox{\rm Im}[(A_2^\ast - B_2^\ast) F_1] \nnb \\
\ek \frac{16}{3} m_B \hat{m}_\ell (1-\hat{s})^2 v 
\mbox{\rm Re}[(A_1^\ast + A_2^\ast + B_1^\ast - B_2^\ast) G_1 +
m_B^2 \hat{s} (A_1^\ast - B_2^\ast) N_1] \nnb \\
\ek \frac{8}{3} m_B (1-\hat{s})^2 v
\mbox{\rm Im}[2 \hat{m}_\ell (A_1^\ast + A_2^\ast - B_1^\ast +
B_2^\ast) G - m_B \hat{s} (A_1^\ast B_1 + A_2^\ast B_2)] \nnb \\
\ar \frac{8}{3} m_B^2 \hat{s} (1-\hat{s})^2 v
\Big( \mbox{\rm Re}[G^\ast N_1 + G_1^\ast N +
m_B^2 \hat{s} N_1^\ast N] + m_B \hat{m}_\ell
\mbox{\rm Im}[(A_2^\ast + B_2^\ast) N] \Big)
\Bigg\}~, \\ \nnb \\
\label{e6822}
P_{TN} \es
\frac{1}{\Delta} \Bigg\{
-16 \hat{s} (1-\hat{s}) v \mbox{\rm Re}[G^\ast H_1 + G_1^\ast H] \nnb \\
\ek 16 m_B \hat{m}_\ell \hat{s} (1-\hat{s}) v 
\mbox{\rm Im}[(A_2^\ast + B_2^\ast) H] \nnb \\
\ar \frac{16}{v^2} f_B \hat{s} \bigg[ 2 v - (1-v^2) 
\ln \bigg( \frac{1+v}{1-v}\bigg) \bigg]
\mbox{\rm Re}[F^\ast G_1] \nnb \\
\ar 2 f_B m_B^3 \hat{m}_\ell (1-\hat{s})^2 v
\mbox{\rm Im}[ A_1^\ast (F_1+F) + B_1^\ast (F_1-F)]
({\cal I}_{8}-{\cal I}_{9}) \nnb \\
\ek 32 \hat{s}^2 v \mbox{\rm Im}[ H_1^\ast H] \nnb \\
\ek f_B^2 m_B^4 \hat{s} \mbox{\rm Im}[ F_1^\ast F]
\Big\{ v [ 3 + v^2 - \hat{s} (1-v^2)] {\cal I}_3 -
(1-v^2) [ (1-\hat{s}) {\cal I}_5 + (1+\hat{s}) {\cal I}_6 ] \Big\} \nnb \\
\ek 4 f_B m_B^4 (1-\hat{s}) [\hat{s} - 2 \hat{m}_\ell^2 (1+\hat{s})] v
\mbox{\rm Im}[ F_1^\ast N] {\cal I}_{8} \nnb \\
\ar 16 f_B \hat{s} \bigg[ 2 v - (1-v^2) \ln \bigg( \frac{1+v}{1-v}\bigg)
\bigg] \mbox{\rm Im}[F_1^\ast G] \nnb \\
\ek 2 f_B m_B^3 \hat{m}_\ell (1-\hat{s}^2) v 
\mbox{\rm Im}[(A_2^\ast+B_2^\ast) F] ({\cal I}_{8}-{\cal I}_{9}) \nnb \\
\ar 4 f_B m_B^2 \hat{s} v \mbox{\rm Re}[F^\ast H_1]  
\Big[ (1-\hat{s}) {\cal I}_{8} + (1+\hat{s}) {\cal I}_{9} \Big] \nnb \\
\ar 4 f_B m_B^2 \hat{s} v \mbox{\rm Im}[F_1^\ast H]
\Big[ (1-\hat{s}) v^2 {\cal I}_{8} + (1-\hat{s} + 2 \hat{s} v^2) 
{\cal I}_{9}\Big] \nnb \\
\ar \frac{16}{v^2} f_B m_B^2 \hat{s} \bigg[2 v - \ln \bigg( \frac{1+v}{1-v}
\bigg) \bigg] \mbox{\rm Re}[F^\ast N_1] \nnb \\
\ar \frac{8}{v^2} f_B m_B \hat{m}_\ell (1-\hat{s})
\bigg[2 v - (1-v^2) \ln \bigg( \frac{1+v}{1-v} \bigg) \bigg]
\mbox{\rm Im}[(A_2^\ast - B_2^\ast) F_1] \nnb \\
\ek \frac{16}{3} m_B \hat{m}_\ell (1-\hat{s})^2 v
\mbox{\rm Re}[(A_1^\ast - A_2^\ast + B_1^\ast + B_2^\ast) G_1 -
m_B^2 \hat{s} (A_1^\ast - B_2^\ast) N_1] \nnb \\
\ar \frac{8}{3} m_B (1-\hat{s})^2 v
\mbox{\rm Im}[2 \hat{m}_\ell (A_1^\ast - A_2^\ast - B_1^\ast -
B_2^\ast) G - m_B \hat{s} (A_1^\ast B_1 + A_2^\ast B_2)] \nnb \\
\ar \frac{8}{3} m_B^2 \hat{s} (1-\hat{s})^2 v    
\Big( \mbox{\rm Re}[G^\ast N_1 + G_1^\ast N +   
m_B^2 \hat{s} N_1^\ast N] + m_B \hat{m}_\ell 
\mbox{\rm Im}[(A_2^\ast + B_2^\ast) N] \Big)
\Bigg\}~, \\ \nnb \\
\label{e6823}
P_{NN} \es
\frac{1}{\Delta} \Bigg\{
- 16 m_B \hat{m}_\ell \hat{s} (1-\hat{s}) 
\mbox{\rm Im}[(A_2^\ast + B_2^\ast) H_1] \nnb \\
\ar 16 \hat{s} v^2 \Big[ (1-\hat{s}) \mbox{\rm Re}[G^\ast H]
+ \hat{s} \vel H \ver^2 \Big] \nnb \\
\ek - 16 \hat{s} (1-\hat{s})  
\mbox{\rm Re}[G_1^\ast H_1 - v^2 G^\ast H] \nnb \\
\ek 16 \hat{s}^2 \Big( \vel H_1 \ver^2 - v^2 \vel H \ver^2 \Big) \nnb \\
\ar f_B^2 m_B^4 \hat{s} \Big[ (1+v^2) {\cal I}_3 - (1-v^2) {\cal I}_6 \Big] 
\vel F \ver^2 \nnb \\
\ek f_B^2 m_B^4 \hat{s} v \Big\{ v [ 2 - \hat{s} (1-v^2) {\cal I}_3    
- (1-v^2) [ (1 - \hat{s}) {\cal I}_5 + \hat{s} v {\cal I}_6 ] \Big\}
\vel F_1 \ver^2 \nnb \\
\ek 4 f_B m_B^2 \hat{s} v^2 \mbox{\rm Re}[F^\ast H]
\Big[ (1-\hat{s}) {\cal I}_{8} + (1+\hat{s}) {\cal I}_{9} \Big] \nnb \\
\ar 4 f_B m_B^2 \hat{s} \mbox{\rm Im}[F_1^\ast H_1] 
\Big[ (1-\hat{s}) v^2 {\cal I}_{8} + (1 - \hat{s} + 
2 \hat{s} v^2) {\cal I}_{9} \Big] \nnb \\
\ar \frac{16}{v} f_B m_B \hat{m}_\ell \hat{s} 
\bigg[ 2 v - (1-v^2) \ln \bigg(\frac{1+v}{1-v} \bigg) \bigg] 
\mbox{\rm Re}[(A_2^\ast + B_2^\ast) F_1] \nnb \\ 
\ar \frac{16}{v} f_B \hat{s} \bigg[ 2 v - (1-v^2) \ln \bigg(
\frac{1+v}{1-v} \bigg) \bigg] \Big( \mbox{\rm Im}[F_1^\ast G_1] -
\mbox{\rm Re}[F^\ast G] \Big) \nnb \\
\ar \frac{8}{v} f_B m_B^2 \hat{s} \bigg[ 4 v - (3 - \hat{s} - v^2 +
\hat{s} v^2) \ln \bigg( \frac{1+v}{1-v} \bigg) \bigg] 
\mbox{\rm Im}[F_1^\ast N_1] \nnb \\
\ek \frac{16}{v} f_B m_B^2 \hat{s} \bigg[ 2 v -                                  
\ln \bigg( \frac{1+v}{1-v} \bigg) \bigg] \mbox{\rm Re}[F^\ast N] \nnb \\
\ar \frac{4}{3} m_B^2 \hat{s} (1-\hat{s})^2 v^2 
\Big( 2 \mbox{\rm Re}[A_1^\ast B_1 + A_2^\ast B_2 - G^\ast N] 
- m_B^2 \hat{s} \vel N \ver^2 \Big)\nnb \\
\ar \frac{4}{3} m_B^2 \hat{s} (1-\hat{s})^2 (3 - 2 v^2) 
\Big( 2 \mbox{\rm Re}[G_1^\ast N_1] + 
m_B^2 \hat{s} \vel N_1 \ver^2 \Big)
\Bigg\}~, \\ \nnb \\
\label{e6824}
P_{TT} \es
\frac{1}{\Delta} \Bigg\{
16 m_B \hat{m}_\ell \hat{s} (1-\hat{s}) 
\mbox{\rm Im}[(A_2^\ast + B_2^\ast) H_1] \nnb \\
\ek 16 \hat{s} v^2 \Big[ (1-\hat{s}) \mbox{\rm Re}[G^\ast H]
+ \hat{s} \vel H \ver^2 \Big] \nnb \\
\ar \frac{1}{2} f_B^2 m_B^4 \Big\{ - (1-\hat{s})^2 (1-v^2) {\cal I}_1
+ [(1-\hat{s})^2 - v^2 + 3 \hat{s} v^2 (2-\hat{s}) +
2 \hat{s}^2 v^4] {\cal I}_3 \nnb \\
\ek (1-v^2) (1-\hat{s})^2 {\cal I}_4 - 2 \hat{s} (1-\hat{s}) v (1-v^2) {\cal I}_5 +
(1-v^2) [1- \hat{s} (2 -\hat{s} + 2 \hat{s} v^2)] {\cal I}_6 \Big\}
\vel F_1 \ver^2 \nnb \\
\ar \frac{1}{2} f_B^2 m_B^4 \Big\{ - (1-\hat{s})^2 (1-v^2) {\cal I}_1
+ [1 -v^2 - 4 \hat{s} + \hat{s}^2 (1-v^2)] {\cal I}_3 \nnb \\
\ek (1-v^2) (1-\hat{s})^2 {\cal I}_4 + (1-v^2) (1-\hat{s}^2) {\cal I}_6 \Big\}
\vel F \ver^2 \nnb \\
\ek 4 f_B m_B^3 \hat{m}_\ell (1-\hat{s})^2 
\mbox{\rm Re}[(A_1^\ast + B_1^\ast) F] ({\cal I}_{8}-{\cal I}_{9}) \nnb \\
\ar 4 f_B m_B^2 \hat{s} v^2
\mbox{\rm Re}[F^\ast H] [(1-\hat{s}) {\cal I}_{8} + (1+\hat{s}) {\cal I}_{9}] \nnb \\
\ek 4 f_B m_B^2 \hat{s}
\mbox{\rm Im}[F_1^\ast H_1] [(1-\hat{s}) v^2 {\cal I}_{8} + 
(1-\hat{s}+2 \hat{s} v^2) {\cal I}_{9}] \nnb \\
\ek f_B m_B \hat{m}_\ell (1-\hat{s}) 
\mbox{\rm Re}[(A_2^\ast + B_2^\ast)F_1]
\Big[ 8 - m_B^2 (4 - v^2 - 4 \hat{s} + 5 \hat{s} v^2) {\cal I}_{8} \nnb \\
\ar m_B^2 (3-3 \hat{s} + 4 \hat{s} v^2) {\cal I}_{9}\Big] \nnb \\
\ar f_B m_B^2 \hat{s}  \mbox{\rm Re}[F^\ast N]
\Big[8 (1+\hat{s}) - m_B^2 (1-\hat{s}) (3-\hat{s}) v^2 {\cal I}_{8} -
m_B^2 (1-\hat{s}^2) {\cal I}_{9} \Big] \nnb \\
\ek f_B m_B^2 \hat{s} \mbox{\rm Im}[F_1^\ast N_1]
\Big[ 8 (1 -\hat{s} + 2 \hat{s} v^2) -
m_B^2 (1-\hat{s}) (4 - v^2 - 4 \hat{s} + 
5 \hat{s} v^2 - 2 \hat{s} v^4) {\cal I}_{8} \nnb \\
\ar m_B^2 (1-\hat{s}) (1-\hat{s} -2 v^2) {\cal I}_{9} \Big] \nnb \\
\ek f_B \mbox{\rm Im}[F_1^\ast G_1]
\Big[ 8 (1 - \hat{s} + 2 \hat{s} v^2) -
m_B^2 (1-\hat{s}) (4 - 5 v^2 -4 \hat{s} + 
9 \hat{s} v^2 -2 \hat{s} v^4) {\cal I}_{8} \nnb \\
\ar m_B^2 (1-\hat{s}) (3-3\hat{s} - 4 v^2 + 
6 \hat{s} v^2) {\cal I}_{9} \Big] \nnb \\
\ar f_B \mbox{\rm Re}[F^\ast G]
\Big[ 8 (1 +\hat{s}) +         
m_B^2 (1-\hat{s}) (4 - 4\hat{s} - 3 v^2 + \hat{s} v^2) {\cal I}_{8} \nnb \\
\ek m_B^2 (1-\hat{s}) (5-3\hat{s} -4 v^2) {\cal I}_{9} \Big] \nnb \\
\ar \frac{16}{3} m_B \hat{m}_\ell (1-\hat{s})^2 
\Big[ 2 \mbox{\rm Im}[(A_2^\ast - B_2^\ast) G_1] -
2 \mbox{\rm Re}[(A_1^\ast + B_1^\ast) G ] \nnb \\
\ar m_B \hat{m}_\ell \Big( \vel A_1 \ver^2 + \vel A_2 \ver^2 +\vel B_1 \ver^2   
+\vel B_1 \ver^2 \Big) \Big] \nnb \\
\ar \frac{64}{3 \hat{s}} \hat{m}_\ell^2 (1-\hat{s})^2 
\Big( \vel G_1 \ver^2 + \vel G \ver^2 \Big) \nnb \\
\ar \frac{8}{3} m_B^2 (1-\hat{s})^2 
\Big( \hat{s} \mbox{\rm Re}[A_1^\ast B_1 + A_2^\ast B_2] 
+ m_B \hat{m}_\ell \mbox{\rm Im}[(A_2^\ast + B_2^\ast) N_1] \Big) \nnb \\
\ar \frac{4}{3} m_B^4 \hat{s}^2 (1-\hat{s})^2
\Big[ (1-2 v^2)  \vel N_1 \ver^2 + v^2 \vel N \ver^2 \Big] \nnb \\
\ar \frac{8}{3} m_B^2 \hat{s} (1-\hat{s})^2
\mbox{\rm Re}[(1-2 v^2) G_1^\ast N_1 + v^2 G^\ast N]
\Bigg\}~, \\ \nnb \\
\mbox{\rm where,} \nnb \\ \nnb \\
\Delta \es
\label{e6825}
16 m_B \hat{m}_\ell (1-\hat{s})^2
\Big( \mbox{\rm Im}[ (A_2^\ast + B_2^\ast) G_1] -
\mbox{\rm Re}[ (A_1^\ast + B_1^\ast) G - m_B \hat{m}_\ell 
(A_1^\ast B_1 + A_2^\ast B_1)] \Big) \nnb \\
\ar 48 m_B \hat{m}_\ell \hat{s} (1-\hat{s}) 
\mbox{\rm Im}[ (A_2^\ast + B_2^\ast) H_1] \nnb \\
\ek 8 m_B^3 \hat{m}_\ell \hat{s} (1-\hat{s})^2 
\mbox{\rm Im}[ (A_2^\ast + B_2^\ast) N_1] \nnb \\
\ar \frac{2}{3} (1-\hat{s})^2                                   
\Big[ 4 (3-v^2) \Big( \vel G_1 \ver^2 + \vel G \ver^2 \Big) +
m_B^2 \hat{s} (3+v^2) \Big( \vel A_1 \ver^2 + \vel A_2 \ver^2 +
\vel B_1 \ver^2 + \vel B_2 \ver^2 \Big) \Big] \nnb \\
\ar 16 \hat{s} v^2 \Big[ (1-\hat{s}) \mbox{\rm Re}[G^\ast H] + 
\hat{s} \vel H \ver^2 \Big] \nnb \\
\ar 16 \hat{s} (3-2 v^2) \Big[ (1-\hat{s}) \mbox{\rm Re}[G_1^\ast H_1] +            
\hat{s} \vel H_1 \ver^2 \Big] \nnb \\
\ek \frac{4}{3} m_B^2 \hat{s} (1-\hat{s})^2 (3-2 v^2) 
\Big( 2 \mbox{\rm Re}[G_1^\ast N_1] + m_B^2 \hat{s} 
\vel N_1 \ver^2 \Big) \nnb \\
\ek \frac{4}{3} m_B^2 \hat{s} (1-\hat{s})^2 v^2 
\Big( 2 \mbox{\rm Re}[G^\ast N] + m_B^2 \hat{s} \vel N \ver^2 \Big) \nnb \\
\ek \frac{1}{2} f_B^2 m_B^4 \vel F \ver^2 
\Big\{ (1-\hat{s})^2 v^2 ({\cal I}_1 + {\cal I}_4) -
(1+\hat{s}^2 + 2 \hat{s} v^2) {\cal I}_3 -
[1-\hat{s} (4 - \hat{s} -2 v^2)] {\cal I}_6 \Big\} \nnb \\
\ar \frac{1}{2} f_B^2 m_B^4 \vel F_1 \ver^2 \Big\{ - (1-\hat{s})^2 v^2 
({\cal I}_1 + {\cal I}_4) +
[1 - \hat{s} (2 - \hat{s} - 4 v^2 + 2 \hat{s} v^2 - 
2 \hat{s} v^4)] {\cal I}_3 \nnb \\ 
\ek 2 \hat{s} (1-\hat{s}) v (1-v^2) {\cal I}_5 +       
[1 - \hat{s} (2 - \hat{s} + 2 \hat{s} v^2 - 2 \hat{s} v^4)] {\cal I}_6 \Big\} \nnb \\
\ek 4 f_B m_B^2 \hat{s} v \mbox{\rm Re} [F^\ast H] 
[(1-\hat{s}) v {\cal I}_{8} + (1+\hat{s}) {\cal I}_{9}] \nnb \\
\ek 4 f_B m_B^2 \hat{s} \mbox{\rm Im} [F_1^\ast H_1]
 [(1-\hat{s}) v^2 {\cal I}_{8} + 
(3 - 2 v^2 - 3 \hat{s} + 4 \hat{s} v^2) {\cal I}_{9}] \nnb \\
\ar 2 f_B m_B \hat{m}_\ell \mbox{\rm Re} [(A_1^\ast + B_1^\ast) F]
\Big[ 8 (1+\hat{s}) +
m_B^2 (1-\hat{s}^2) v^2 {\cal I}_{8} +
m_B^2 (1-\hat{s}) (1-3\hat{s}) {\cal I}_{9} \Big] \nnb \\
\ek f_B m_B \hat{m}_\ell (1-\hat{s}) 
\mbox{\rm Re}[(A_2^\ast + B_2^\ast) F_1] 
\Big[ 8 + m_B^2 (1 - 5 \hat{s}) v^2 {\cal I}_{8} +
m_B^2 (3 - 3\hat{s} + 4 \hat{s} v^2) {\cal I}_{9} \Big] \nnb \\
\ar f_B \mbox{\rm Im}[F_1^\ast G_1]
\Big[ -24 (1-\hat{s} + 2 \hat{s} v^2) +
m_B^2 (1-\hat{s}) (1+3 \hat{s} - 6 \hat{s} v^2) v^2 {\cal I}_{8} \nnb \\
\ar m_B^2 (1-\hat{s}) (1-\hat{s} - 2 \hat{s} v^2) {\cal I}_{9} \Big] \nnb \\
\ar f_B \mbox{\rm Re}[F^\ast G]  
\Big[ -24 (1+\hat{s}) +
m_B^2 (1-\hat{s}) (1-3 \hat{s}) v^2 {\cal I}_{8} -
m_B^2 (1-\hat{s}) (1-7 \hat{s} + 4 \hat{s} v^2) {\cal I}_{9} \Big] \nnb \\
\ar f_B m_B^2 \hat{s} \mbox{\rm Im}[F_1^\ast N_1]
\Big[ -8 (1-\hat{s} + 2 \hat{s} v^2) +
m_B^2 (1-\hat{s}) (3 + \hat{s} - 2 \hat{s} v^2) v^2 {\cal I}_{8} \nnb \\
\ar m_B^2 (1-\hat{s}) (3 - 2 v^2 - 3 \hat{s} + 4 \hat{s} v^2 ) 
{\cal I}_{9} \Big] \nnb \\
\ar f_B m_B^2 \hat{s} \mbox{\rm Re}[F^\ast N]  
\Big[ -8 (1+\hat{s}) +
m_B^2 (1-\hat{s}) (3 - \hat{s}) v^2 {\cal I}_{8} +
m_B^2 (1-\hat{s}^2) {\cal I}_{9} \Big]~.
\eea
In Eqs. (\ref{e6816})--(\ref{e6825}), $\hat{s} = q^2/m_B^2$, 
$v=\sqrt{1-4 \hat{m}_\ell^2/\hat{s}}$ is the lepton
velocity with $\hat{m}_\ell = m_\ell/m_B$, and 
${\cal I}_i$ represent the following integrals
\bea
{\cal I}_i = \int_{-1}^{+1} {\cal F}_i(z) dz~,\nnb
\eea
where
\bea
\begin{array}{lll}
{\cal F}_{1}  = \ds\frac{z^2}{(p_1 \cdot k) (p_2 \cdot k)}~,&    
{\cal F}_{2}  = \ds\frac{z}{(p_1 \cdot k) (p_2 \cdot k)}~,&    
{\cal F}_{3}  = \ds\frac{1}{(p_1 \cdot k) (p_2 \cdot k)}~, \\ \\  
{\cal F}_{4}  = \ds\frac{z^2}{(p_1 \cdot k)^2}~,&    
{\cal F}_{5}  = \ds\frac{z}{(p_1 \cdot k)^2}~,&    
{\cal F}_{6}  = \ds\frac{1}{(p_1 \cdot k)^2}~, \\ \\
{\cal F}_{7}  = \ds\frac{z}{(p_2 \cdot k)^2}~,&                           
{\cal F}_{8} = \ds\frac{z^2}{p_1 \cdot k}~,&         
{\cal F}_{9} = \ds\frac{1}{p_1 \cdot k}~.
\end{array} \nnb
\eea

\section{Numerical analysis and discussion}

We now proceed by presenting our numerical analysis for all possible
double--lepton polarizations. The values of the input parameters 
which have been used in the present work are:
$\vel V_{tb} V_{ts}^\ast \ver = 0.0385$, $m_\mu=0.106~GeV$, $m_\tau=1.78~GeV$, 
$m_b=4.8~GeV$. For the SM values of the Wilson coefficients
we have used $C_7^{SM}(m_b)=-0.313$, $C_9^{SM}(m_b) = 4.344$ and
$C_{10}^{SM}(m_b) = -4.669$. The magnitude of $C_7^{SM}$ is quite well constrained
from the $b \rar s \gamma$ transition, and hence it is well established.
Therefore the values of $C_{BR}$ and $C_{SL}$ are fixed by the relations
$C_{BR}=-2 m_b C_7^{eff}$ and $C_{SL}=-2 m_s C_7^{eff}$. It is well known
that the Wilson coefficient $C_9^{SM}$ receives also long distance
contributions which have their origin in the real $\bar{c}c$ intermediate
states, i.e., $J/\psi$, $\psi^\prime$, $\cdots$ \cite{R6814}. In the present
work we restrict ourselves only to short distance contributions.

In performing the numerical analysis, as is obvious from the expressions of
$P_{ij}$ given in Eqs. (\ref{e6816})--(\ref{e6824}), we need to know the
values of the new Wilson coefficients. During the numerical calculations, we
will vary all new Wilson coefficients in the range $- \vel C_{10}^{SM} \ver
\le C_X \le \vel C_{10}^{SM} \ver$ and assume that all new Wilson
coefficients are real. The experimental results on the branching ratio of the
$B \rar K^\ast (K) \ell^+ \ell^-$ decays  \cite{R6815,R6816} and the bound 
on the branching ratio of $B \rar \mu^+ \mu^-$ \cite{R6817} suggest that
this is the right order of magnitude for the Wilson coefficients describing
the vector and scalar interaction coefficients. But present experimental
results on the branching ratio of the 
$B \rar K^\ast \ell^+ \ell^-$ and $B \rar K \ell^+ \ell^-$ decays impose
stronger restrictions on some of the new Wilson coefficients. For example,
$-2 \le C_{LL} \le 0$, $0 \le C_{RL} \le 2.3$, $-1.5 \le C_{T} \le 1.5$ and
$-3.3 \le C_{TE} \le 2.6$, and all of the remaining Wilson coefficients vary
in the region $- \vel C_{10}^{SM} \ver \le C_X \le \vel C_{10}^{SM} \ver$.

In further numerical analysis, as can easily be seen from Eqs.
(\ref{e6816})--(\ref{e6824}), explicit forms of the form factors are needed,
which are the main and most important parameters in the calculation of
$P_{ij}$. These form factors are calculated in the framework of the QCD sum
rules in \cite{R6803,R6812,R6813} and their $q^2$ dependence can be
represented, to a very good accuracy, in the following forms
\bea
\begin{array}{ll}
g_(q^2) = \ds\frac{1~GeV}{\ds \ga 1-\frac{q^2}{(5.6~GeV)^2}\dr^2}~,&
f_(q^2) = \ds\frac{0.8~GeV}{\ds \ga 1-\frac{q^2}{(6.5~GeV)^2}\dr^2}~,\\ \\  
g_1(q^2) = \ds\frac{3.74~GeV^2}{\ds \ga 1-\frac{q^2}{40.5~GeV^2}\dr^2}~,&
f_1(q^2) = \ds\frac{0.68~GeV^2}{\ds \ga 1-\frac{q^2}{30~GeV^2}\dr^2}~,
\end{array} \nnb
\eea
which we will be using in the numerical calculations.

Numerical results
are presented only for the $B_s \rar \ell^+ \ell^- \gamma$ decay. It is
clear that in the SU(3) limit the difference between the decay rates is
attributed to
the CKM matrix elements only, i.e.,
\bea
\frac{\Gamma(B_d \rar \ell^+ \ell^- \gamma)}
{\Gamma(B_s \rar \ell^+ \ell^- \gamma)} \simeq \frac{\vel V_{tb}
V_{td}^\ast\ver^2}
{\vel V_{tb} V_{ts}^\ast\ver^2} \simeq \frac{1}{20}~.\nnb
\eea 

It follows from the explicit expressions of the double--lepton polarization
asymmetries that they depend on $q^2$ and the new Wilson coefficients. For
this reason there may appear difficulties in studying the dependencies of
the physical properties on both parameters at the same time. Hence, it is
necessary to eliminate the dependence of $P_{ij}$ on one of these
parameters. Here in the present work, we eliminate $q^2$ dependence of
$P_{ij}$ by performing integration over $q^2$ in the kinematically allowed
region. The averaging of $P_{ij}$ over $q^2$ is defined as
\bea
\label{e6826}
\lla P_{ij} \rra = \ds \frac{\int_{4 m_\ell^2}^{m_B^2}
P_{ij} \ds \frac{d {\cal B}}{dq^2} dq^2}
{\int_{4 m_\ell^2}^{m_B^2}
\ds \frac{d {\cal B}}{dq^2} dq^2}
\eea
The reason why we study the dependence of $\lla P_{ij} \rra$ on new Wilson
coefficients is that in doing so we directly establish new physics beyond
the SM, if the value of $\lla P_{ij} \rra$ turns out to be different
compared to that predicted by the SM. 

In Figs. (1)--(8) we present the dependence of the averaged double--lepton
polarization asymmetries on the new Wilson coefficients. From these figures
we get the following results:

\begin{itemize}

\item $\lla P_{LL} \rra$ exhibits very strong dependence on scalar, tensor
interactions as well as on the vector interaction with coefficient $C_{RL}$.
When scalar interaction coefficients vary in the region $-4 \le C_{scalar}
\le -0.8$ the value of $\lla P_{LL} \rra$ is positive; in the region $-0.8\le
C_{scalar} \le 0.8$ it gets negative value and when it varies in the region 
$0.8\le C_{scalar} \le 4$ it again gets positive values. We should remind
that in the SM $\lla P_{LL} \rra$ is negative and its magnitude is 
$\vel \lla P_{LL} \rra \ver \approx 0.7$. Similar situation holds for tensor
interaction with the coefficient $C_{TE}$, as can easily be seen in Fig. (1).
Our analysis shows that $\lla P_{LT} \rra$ depends more strongly on scalar
and tensor interactions, as is the case for $\lla P_{LL} \rra$. Therefore we
can conclude that measurement of the sign and magnitude of $\lla P_{LL}
\rra$ and $\lla P_{LT} \rra$ can give essential information about the
existence of new physics beyond the SM. 

\item $\lla P_{TL} \rra$ is very sensitive to the existence of the scaler
interaction only. More essential than that, $\lla P_{TL} \rra$ changes its
sign when scalar interaction coefficients vary in the allowed region. Such
behavior can serve as a good test for establishing new type of scalar
interaction.

\item $\lla P_{TT} \rra \approx -\lla P_{NN} \rra$, and both are quite
sensitive to the existence of tensor and scalar interactions, and to the
vector interaction with coefficient $C_{LR}$. In the presence of tensor,
scalar and vector interactions the values of $\lla P_{TT} \rra$ and $\lla
P_{NN} \rra$ can exceed the SM results 6, 5 and 2.5 times, respectively.
Moreover, when scalar interaction coefficients $C_{RLLR} (C_{RLRL})$ is
negative, then the value of $\lla P_{TT} \rra$ is positive (negative). When
$C_{RLLR} (C_{RLRL})$ is positive, $\lla P_{TT} \rra$ becomes negative 
(positive). For the case when vector interaction coefficient $C_{LR}$ is
negative, $\lla P_{TT} \rra$ is also negative, while, when $C_{LR}$ changes
sign and becomes positive, $\lla P_{TT} \rra$ turns out to be positive as
well. Therefore determination of the sign and magnitude of $\lla P_{NN}
\rra$ and $\lla P_{TT} \rra$ can give unambiguous information about the
existence of scalar and vector interactions. Departure of $\lla P_{NT} \rra$
and $\lla P_{TN} \rra$ from SM results is of not considerable importance,
and hence we do not present their dependence on $C_X$.  

\end{itemize}

Depicted in Figs. (4)--(7) are the dependence of $\lla P_{ij} \rra$ on
the new Wilson coefficients for the $B_s \rar \tau^+ \tau^- \gamma$ decay.
Similar to the $B_s \rar \mu^+ \mu^- \gamma$ decay, we observe that several
of the double--lepton polarization asymmetries are very sensitive to the
existence of new physics. More precisely, we can briefly summarize the
results as follows:

i) except $C_{LL}$, $\lla P_{LL} \rra$ is very sensitive to the existence of
all new Wilson coefficients.

ii) $\lla P_{LT} \rra$ and $\lla P_{TL} \rra$ exhibit strong dependence on
all Wilson coefficients, except the vector interactions $C_{LL}$ and
$C_{RR}$. These quantities show very strong dependence, especially,  on all
types of scalar interactions, vector interactions with coefficients $C_{LR}$
and $C_{RL}$, and tensor interaction with coefficient $C_T$.

iii)  $\lla P_{TT} \rra = -\lla P_{NN} \rra$ are very sensitive to the
existence of scalar interactions $C_{RLLR}$ and $C_{LRRL}$, when these
coefficients are both positive. More important than that, $\lla P_{TT} \rra
= -\lla P_{NN} \rra$ change their sign when scalar coefficients vary in the
region $-3 > C_{scalar} > +3$.  

iv) $\lla P_{NT} \rra \approx \lla P_{TN} \rra$ are both strongly 
dependent only on the tensor interaction with coefficient $C_{TE}$. For all
other new Wilson coefficients the values of $\lla P_{NT} \rra \approx
\lla P_{TN} \rra$ are very close to the SM prediction, i.e., to zero.
Furthermore, when $C_{TE}$ is negative (positive), these quantities get
positive (negative) values, with a considerable departure from the SM about
$15\%$. Therefore, determination of the sign and and magnitude of $\lla
P_{TN} \rra$ and $\lla P_{NN} \rra$ can give direct information solely 
about the existence of the tensor interaction. 

v) Similar situation holds for the double--lepton polarization asymmetry 
$\lla P_{LN} \rra$ as well, i.e., $\lla P_{LN} \rra$ shows strong
dependence only on the tensor interaction with coefficient $C_{TE}$. When
$C_{TE}$ is negative (positive), $\lla P_{LN} \rra$ gets negative (positive)
values. Hence, departure from the SM prediction (in SM 
$\lla P_{LN} \rra\approx 0$) can reach to $4\%$. Therefore analysis of $\lla
P_{LN} \rra$ can serve as a good test for establishing the presence of
tensor interactions. The values of $\lla P_{NL} \rra$ is negligibly small
(maximum departure from the SM is being about $1.5\%$) and for this reason
we do not present its dependence on $C_X$.

At the end of this section, let us discuss the possibility of measurement of the
lepton polarization asymmetries in experiments. Experimentally, to measure
an asymmetry $\la P_{ij} \ra$ of the decay with the branching ratio ${\cal B}$
at $n \sigma$ level, the required number of events
(i.e., the number of $B \bar{B}$ pair) are determined by the following
expression
\bea
{\cal N} = \frac{n^2}{{\cal B} s_1 s_2 \la P_{ij} \ra^2}~,\nnb
\eea
where $s_1$ and $s_2$ are the efficiencies of the leptons. Efficiency of 
the $\mu$ lepton is practically equal to one, and typical values of the
efficiency of the $\tau$ lepton ranges from $50\%$ to $90\%$ for
the various decay modes \cite{R6818}.

From the expression for ${\cal N}$ we see that, in order to obtain the 
double--lepton polarization asymmetries in $B_s \rar \ell^+ \ell^- \gamma$
decays at $3\sigma$ level, the minimum number of required events
are (for the efficiency of $\tau$--lepton we take $0.5$):

\begin{itemize}
\item for the $B_s \rar \mu^+ \mu^- \gamma$ decay
\bea
{\cal N} = \left\{ \begin{array}{ll}
\sim 10^{9}     & \lla P_{LL} \rra~,\\
\sim 7 \times 10^{10}  & \lla P_{TT} \rra~, \\
\sim 3 \times10^{10}  & \lla P_{NN} \rra \simeq \lla P_{LT} \rra
\simeq \lla P_{TL} \rra~,
\end{array} \right. \nnb
\eea
which yields that, for detecting $\lla P_{LN} \rra$, $\lla P_{NL} \rra$,
$\lla P_{NT} \rra$  and $\lla P_{TN} \rra$, more than $10^{13}$ $\bar{B} B$
pairs are required.

\item for $B_s \rar \tau^+ \tau^- \gamma$ decay
\bea
{\cal N} = \left\{ \begin{array}{ll}
\sim10^{10}  & \lla P_{LL}\rra,
\lla P_{TT} \rra,~\lla P_{NN} \rra~,\\  
\sim 3\times 10^{11}  & \lla P_{LT} \rra = \lla P_{TL} \rra ~,\\
> 10^{13} & \lla P_{LN} \rra, \lla P_{NL} \rra,
\lla P_{TN} \rra, \lla P_{NT} \rra~.
\end{array} \right.
\nnb
\eea
\end{itemize}

The number of $\bar{B} B$ pairs that will be produced at LHC is
around $\sim 10^{12}$. As a result of a comparison of this number of
$\bar{B} B$ pairs with that of ${\cal N}$, we conclude that $\lla P_{LL}
\rra$, $\lla P_{TT} \rra$, $\lla P_{NN} \rra$, $\lla P_{TL} \rra$ and $\lla
P_{LT} \rra$ in  $B_s \rar \mu^+ \mu^- \gamma$ decay, and $\lla P_{LL}\rra$,
$\lla P_{TT} \rra$ and  $\lla P_{NN} \rra$ in $B_s \rar \tau^+ \tau^- \gamma$
decay can be detectable in future experiments at LHC.
Note that for calculation of the branching ratio, we take its SM result,
i.e., ${\cal B}(B_s \rar \mu^+ \mu^- \gamma) \simeq 1.3 \times 10^{8}$ and
${\cal B}(B_s \rar \tau^+ \tau^- \gamma) \simeq 6 \times 10^{9}$. In
obtaining these values, minimal value for the photon energy is taken to be
$50~MeV$. 

In conclusion, we calculate nine double--lepton polarization asymmetries
using the most general, model independent form of the effective Hamiltonian
including all possible form of interactions. The sensitivity of the
averaged double--lepton polarization asymmetries to the new Wilson
coefficients are studied. Finally we discuss the possibility of experimental
measurement of these double--lepton polarization asymmetries at LHC.

\newpage

\section*{Figure captions}
{\bf Fig. (1)} The dependence of the averaged double--lepton polarization
asymmetry $\lla P_{LL} \rra$ on the new Wilson coefficients for the
$B_s \rar \mu^+ \mu^- \gamma$ decay.\\ \\
{\bf Fig. (2)} The same as in Fig. (1), but for the averaged 
double--lepton polarization asymmetry $\la P_{TL} \ra$.\\ \\
{\bf Fig. (3)} The same as in Fig. (1), but for the averaged
double--lepton polarization asymmetry $\la P_{TT} \ra$.\\ \\
{\bf Fig. (4)} The same as in Fig. (1), but for the
$B_s \rar \tau^+ \tau^- \gamma$ decay.\\ \\
{\bf Fig. (5)} The same as in Fig. (4), but for the averaged
double--lepton polarization asymmetry $\la P_{LT} \ra$.\\ \\
{\bf Fig. (6)} The same as in Fig. (4), but for the averaged
double--lepton polarization asymmetry $\la P_{TL} \ra$.\\ \\
{\bf Fig. (7)} The same as in Fig. (4), but for the averaged
double--lepton polarization asymmetry $\la P_{TT} \ra$.

\newpage

\newpage

\begin{figure}
\vskip 1.5 cm
    \includegraphics{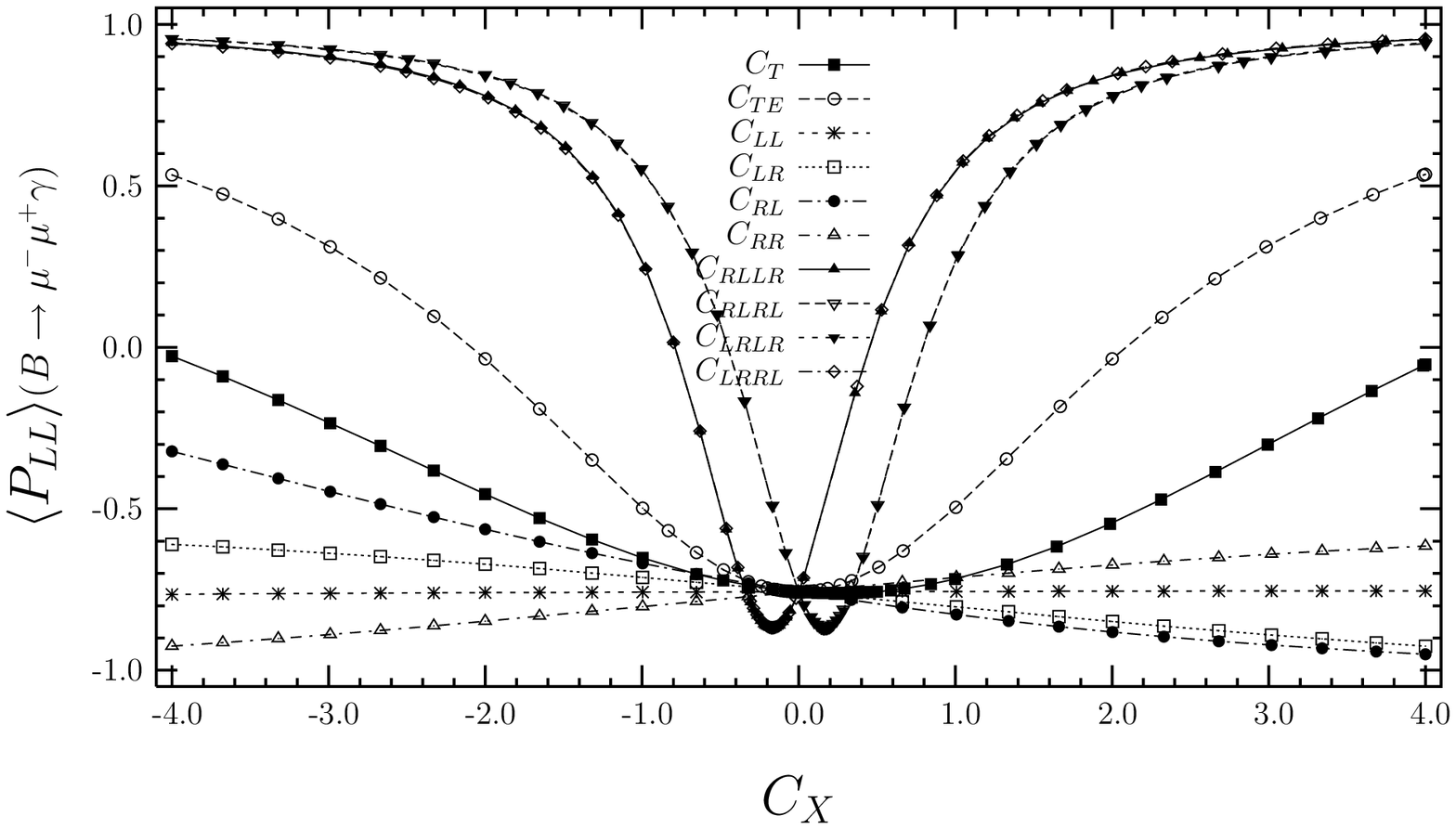}
\vskip 7.8cm
\caption{}
\end{figure}

\begin{figure}
\vskip 2.5 cm
    \includegraphics{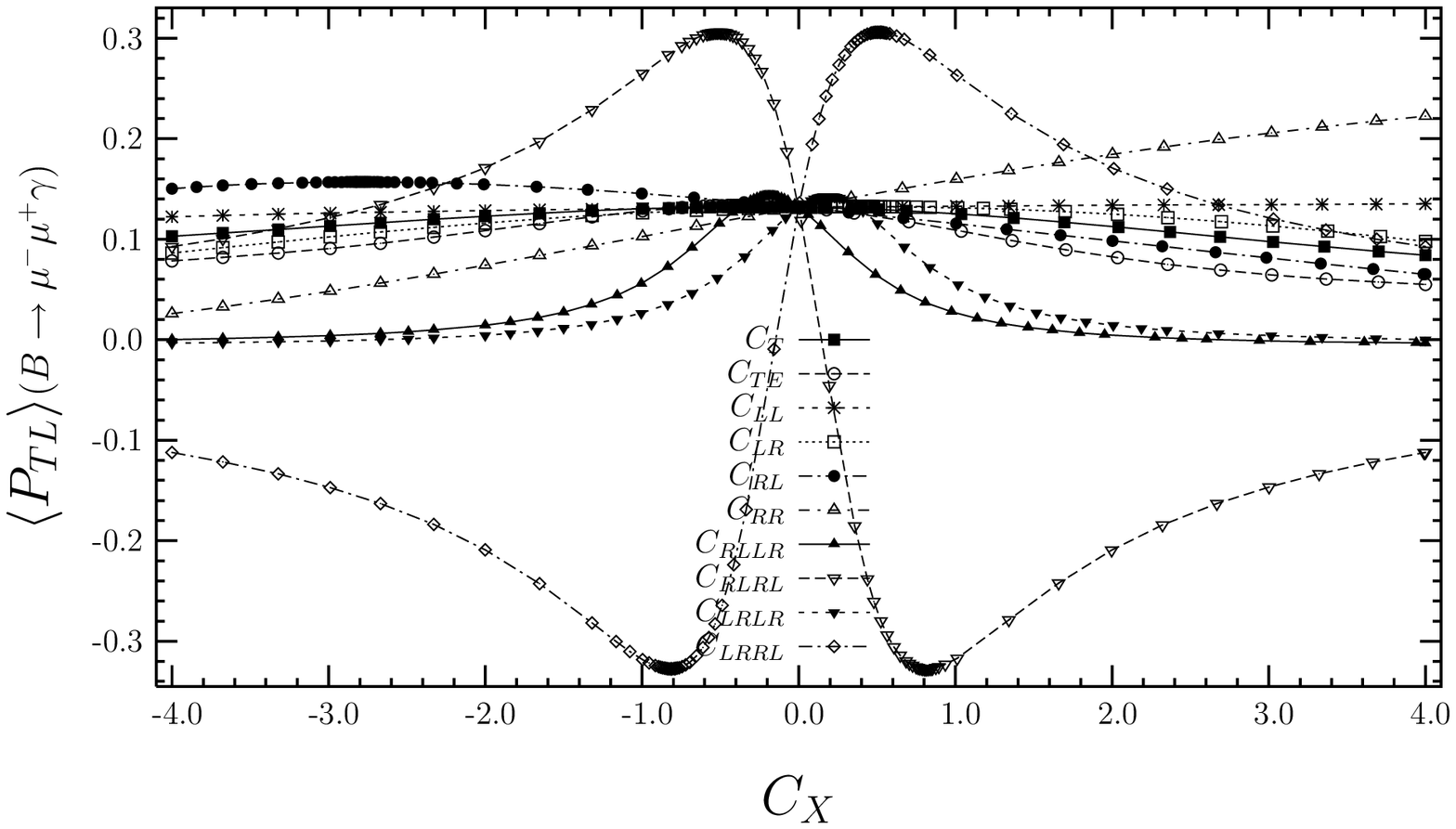}
\vskip 7.8 cm
\caption{}
\end{figure}

\begin{figure}
\vskip 1.5 cm
    \includegraphics{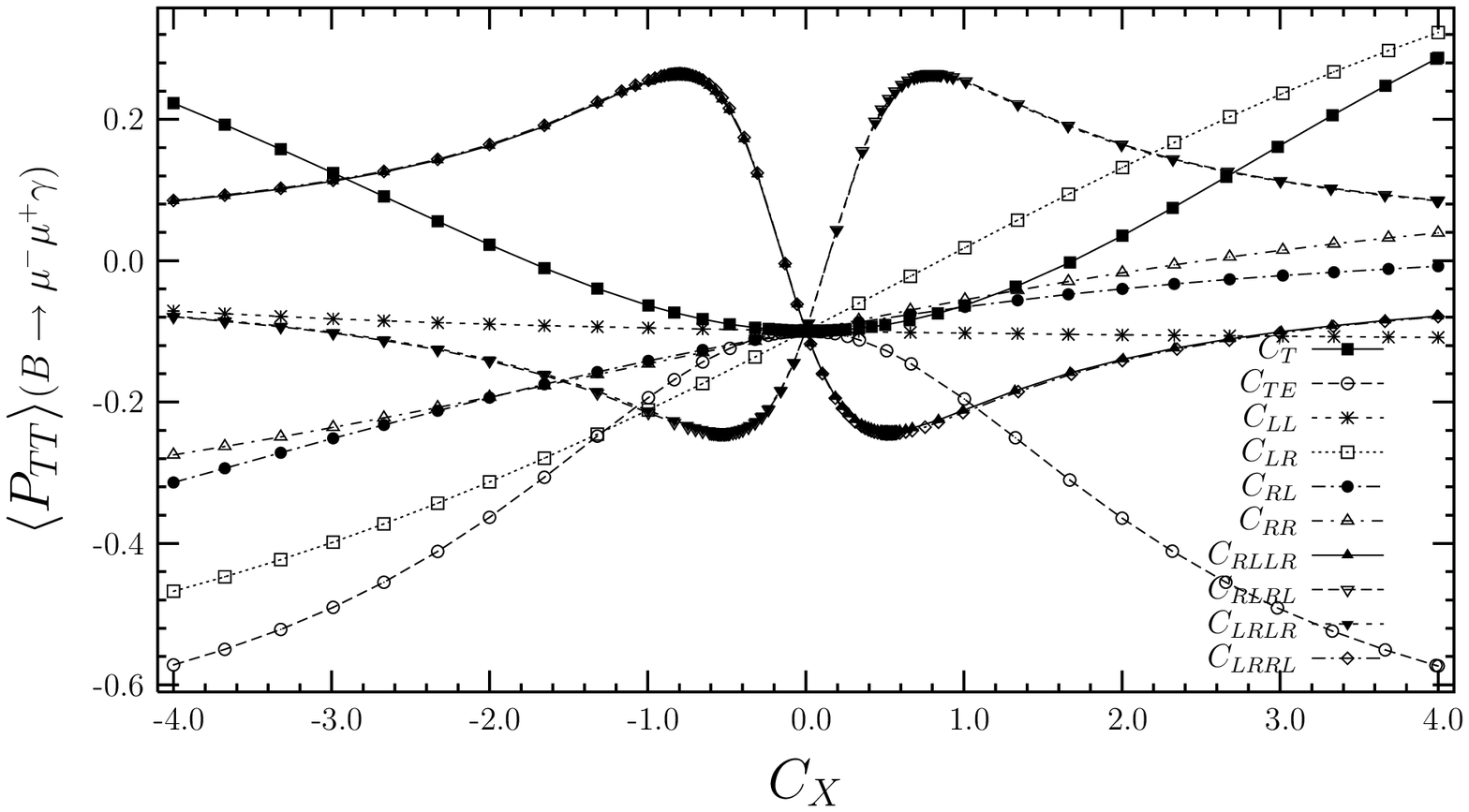}
\vskip 7.8cm
\caption{}
\end{figure}

\begin{figure}
\vskip 2.5 cm
    \includegraphics{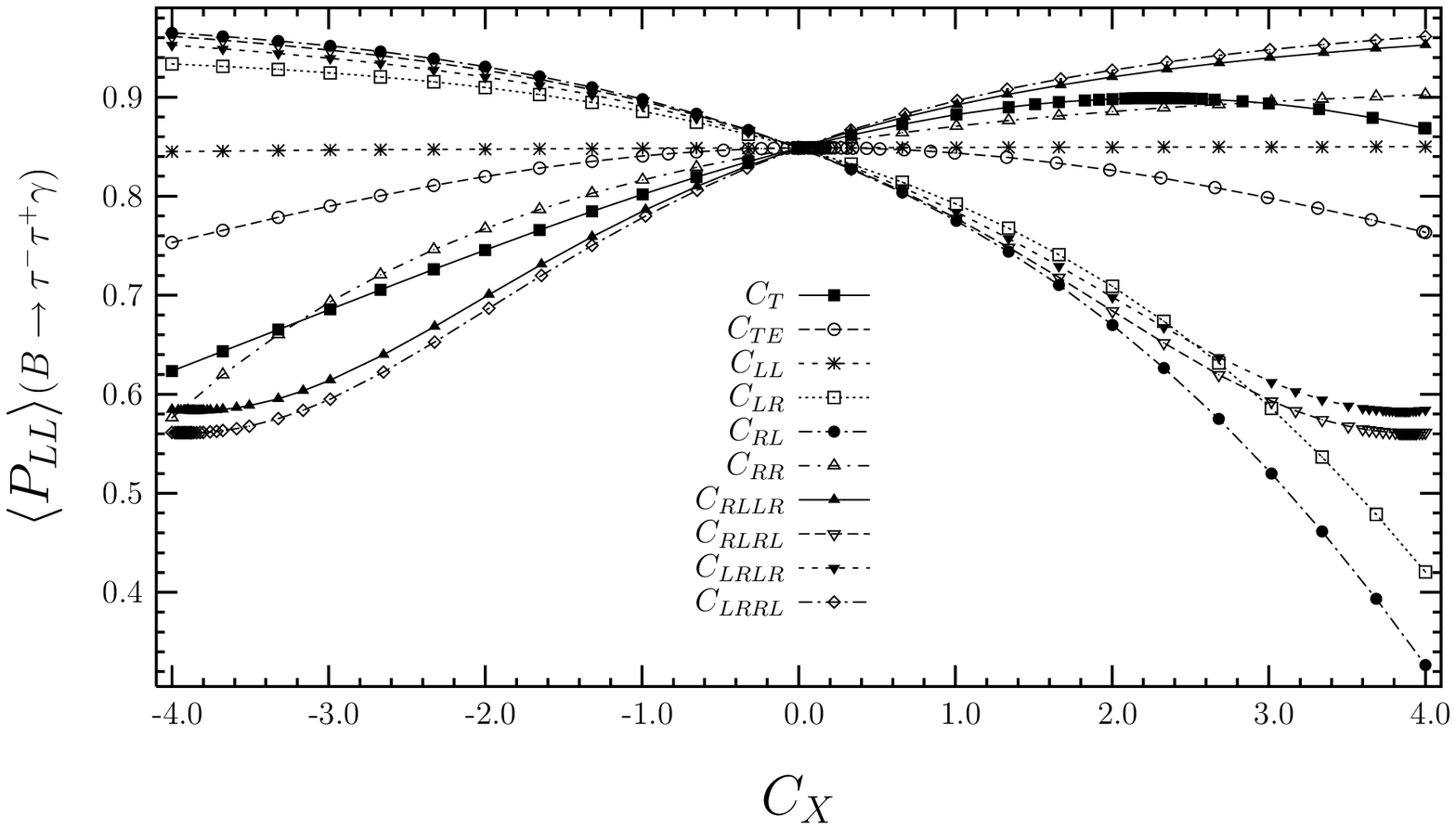}
\vskip 7.8 cm
\caption{}
\end{figure}

\begin{figure}
\vskip 2.5 cm
    \includegraphics{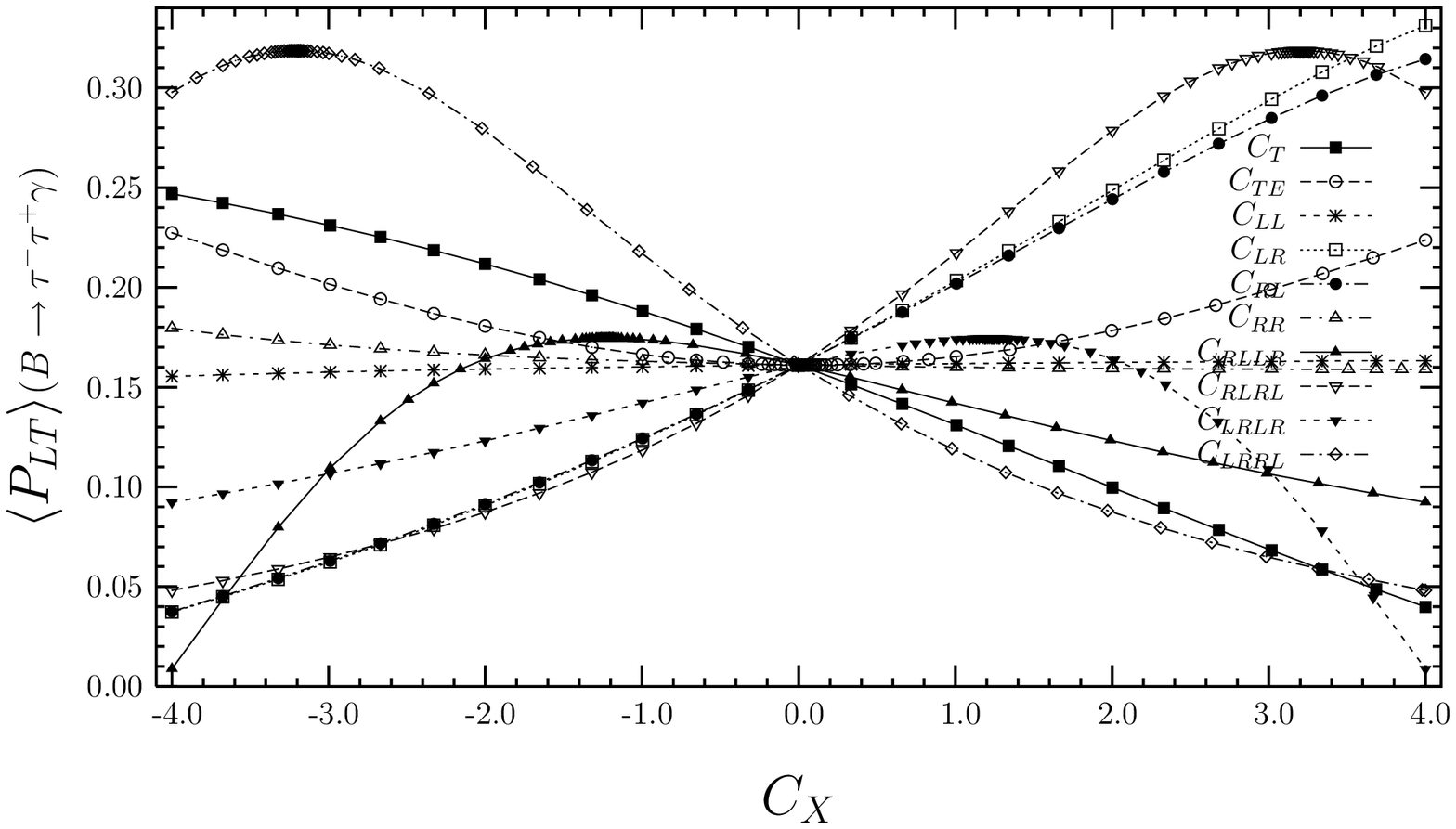}
\vskip 7.8 cm
\caption{}
\end{figure}

\begin{figure}
\vskip 1.5 cm
    \includegraphics{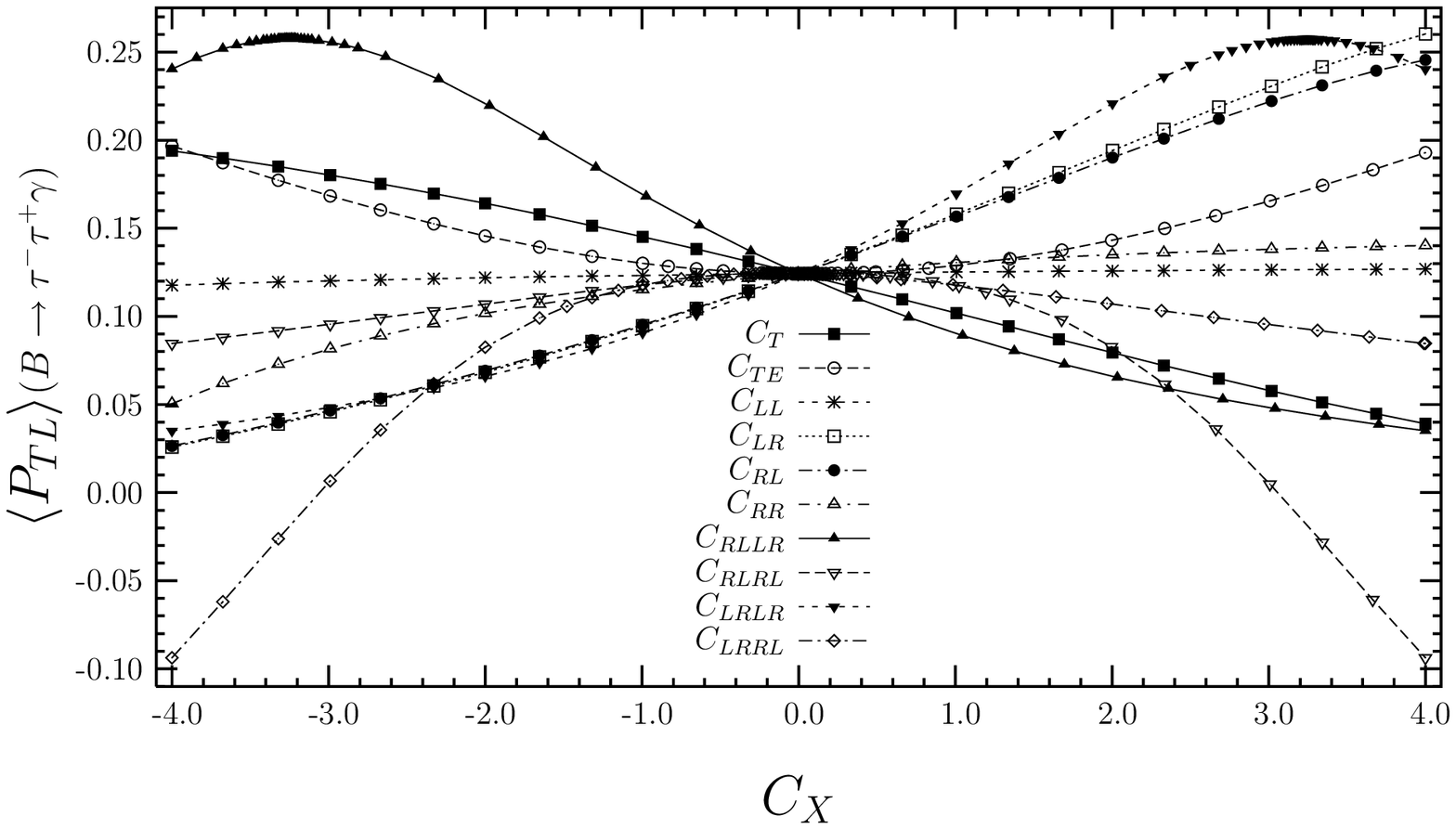}
\vskip 7.8cm
\caption{}
\end{figure}

\begin{figure}
\vskip 2.5 cm
    \includegraphics{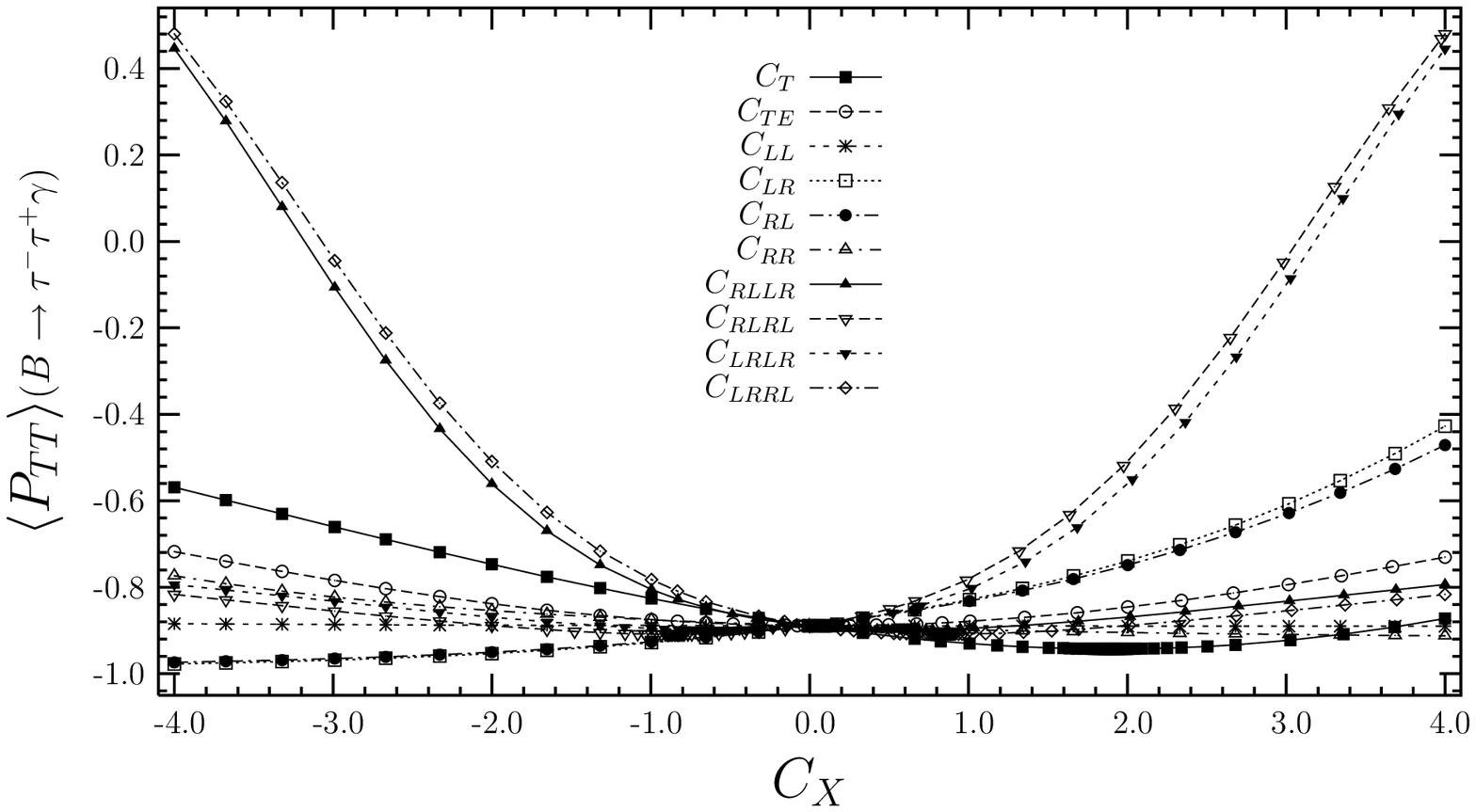}
\vskip 7.8 cm
\caption{}
\end{figure}

\end{document}